\documentclass[twocolumn]{aastex63}

\def\cii{\hbox{{\rm [C {\scriptsize II}]}}}
\def\nii{\hbox{{\rm [N {\scriptsize II}]}}}
\def\13cii{\hbox{{\rm [$^{13}$C {\scriptsize II}]}}}

\def\ci{\hbox{{\rm [C {\scriptsize I}]}}}

\def\oi{\hbox{{\rm [O {\scriptsize I}]}}}
\def\hii{\hbox{{\rm H {\scriptsize II}}}}

\defcitealias{MSP1991}{MSP}
\defcitealias{TFT2007}{TFT07}
\defcitealias{VA2013}{VA13}
\defcitealias{VPHAS_Wd2_2015}{VPHAS+}

\usepackage{multirow}
\usepackage{hyperref}
\usepackage{gensymb}
\usepackage[flushleft]{threeparttable}
\usepackage{amsmath}
\usepackage{changepage}
\usepackage{rotating}
\usepackage{array}
\usepackage{multirow}

\received{June 30, 2023}
\accepted{October 03, 2023}

\shorttitle{GMM}
\shortauthors{Tiwari et al.}

\graphicspath{{./}{figures/}}

\begin{document}

\title{Identifying physical structures in our Galaxy with Gaussian Mixture Models: An unsupervised machine learning technique}

\author{M. Tiwari$^*$}
\affiliation{University of Maryland, Department of Astronomy, College Park, MD 20742-2421, USA}
\affiliation{Max Planck Institute for Radioastronomy, Auf dem H\"{u}gel, 53121 Bonn, Germany}

\author{R. Kievit$^*$}
\affiliation{Leiden Observatory, Leiden University, PO Box 9513, 2300 RA Leiden, The Netherlands}

\author{S. Kabanovic}
\affiliation{I. Physik. Institut, University of Cologne, Z\"{u}lpicher Str. 77, 50937 Cologne, Germany}

\author{L. Bonne}
\affiliation{SOFIA Science Center, USRA, NASA Ames Research Center, M.S. N232-12, Moffett Field, CA 94035, USA}

\author{F. Falasca}
\affiliation{Courant Institute of Mathematical Sciences, New York University, New York, NY, USA}

\author{C. Guevara}
\affiliation{I. Physik. Institut, University of Cologne, Z\"{u}lpicher Str. 77, 50937 Cologne, Germany}

\author{R. Higgins}
\affiliation{I. Physik. Institut, University of Cologne, Z\"{u}lpicher Str. 77, 50937 Cologne, Germany}

\author{M. Justen}
\affiliation{I. Physik. Institut, University of Cologne, Z\"{u}lpicher Str. 77, 50937 Cologne, Germany}

\author{R. Karim}
\affiliation{University of Maryland, Department of Astronomy, College Park, MD 20742-2421, USA}

\author{\"{U}. Kavak}
\affiliation{SOFIA Science Center, USRA, NASA Ames Research Center, M.S. N232-12, Moffett Field, CA 94035, USA}

\author{C. Pabst}
\affiliation{Instituto de F\'isica Fundamental, CSIC, Calle Serrano 121-123, 28006 Madrid, Spain}
\affiliation{Leiden Observatory, Leiden University, PO Box 9513, 2300 RA Leiden, The Netherlands}

\author{M. W. Pound}
\affiliation{University of Maryland, Department of Astronomy, College Park, MD 20742-2421, USA}

\author{N. Schneider}
\affiliation{I. Physik. Institut, University of Cologne, Z\"{u}lpicher Str. 77, 50937 Cologne, Germany}

\author{R. Simon}
\affiliation{I. Physik. Institut, University of Cologne, Z\"{u}lpicher Str. 77, 50937 Cologne, Germany}

\author{J. Stutzki}
\affiliation{I. Physik. Institut, University of Cologne, Z\"{u}lpicher Str. 77, 50937 Cologne, Germany}

\author{M. Wolfire}
\affiliation{University of Maryland, Department of Astronomy, College Park, MD 20742-2421, USA}

\author{A. G. G. M. Tielens}
\affiliation{University of Maryland, Department of Astronomy, College Park, MD 20742-2421, USA}
\affiliation{Leiden Observatory, Leiden University, PO Box 9513, 2300 RA Leiden, The Netherlands}


\def\thefootnote{*}\footnotetext{Both authors contributed equally to this work.}\def\thefootnote{\arabic{footnote}}

\begin{abstract}
We explore the potential of the Gaussian Mixture Model (GMM), an unsupervised machine learning method, to identify coherent physical structures in the ISM. The implementation we present can be used on any kind of spatially and spectrally resolved data set. We provide a step-by-step guide to use these models on different sources and data sets. Following the guide, we run the models on NGC~1977, RCW~120 and RCW~49 using the \cii\,~158~$\mu$m mapping observations from the SOFIA telescope. We find that the models identified 6, 4 and 5 velocity coherent physical structures in NGC~1977, RCW~120 and RCW~49, respectively, which are validated by analysing the observed spectra towards these structures and by comparison to earlier findings. In this work we demonstrate that GMM is a powerful tool that can better automate the process of spatial and spectral analysis to interpret mapping observations.

\end{abstract}


\section{Introduction} \label{sec:intro}

Massive stars ($>$ 8~\(\textup{M}_\odot\)) play one of the most important roles in regulating the physics and chemistry of the interstellar medium (ISM). During their lifetime ($<$ 5~Myr), massive stars inject enormous amounts of radiative energy through extreme-ultraviolet (EUV) and far-UV (FUV) photons \citep{Hollenbach1999}, and mechanical energy through stellar winds \citep{Castor1975,Weaver1977,Pabst2019} into their surroundings. This causes diffuse and dense gas to form different structures either by disrupting or compressing molecular clouds in the vicinity of massive stars \citep{Elmegreen1977,Walborn2002}. These structures can be spatially and spectrally distinct. Their physical conditions are among others dependent on morphology, relative location to the main ionising source and the star formation history of the region \citep{Tiwari2022}. Estimations of these physical conditions quantify the role of stellar feedback in the evolution of the ISM. However, the identification of coherent physical structures in the ISM is not straightforward due to its turbulent nature. Recent observational studies have shown that besides high angular resolution data, a detailed spectral and spatial analysis is necessary to identify coherent physical structures in our Galaxy (e.g., \citealt{Hacar2013,Henshaw2019}).  \\

Soon balloon missions such as ASTROS (Astrophysics Stratospheric Telescope for High Spectral Resolution Observations at Submillimeter-wavelengths, \citealt{Pineda2022}) and GUSTO (Galactic/ Extragalactic Spectroscopic Terahertz Observatory, \citealt{Gusto2022}) will carry out large-area surveys of \nii\,, \cii\ and \oi\ fine-structure lines throughout the Galaxy with high spatial and spectral resolutions.
Besides these balloon missions, the GEco (Galactic Ecology, \citealt{Simon2023}) project is an upcoming large scale survey with the CCAT observatory which will observe CO and \ci\ lines. 
Identifying structures which are spatially and spectrally distinct in these large areas is a big challenge. Manually analysing these huge data sets will be time consuming and cost ineffective. Therefore we emphasize the need of an automated technique to assist in this process.

To this end, we explore the usability of the Gaussian Mixture Model (GMM), an automated technique to identify distinct physical structures in three Galactic sources: NGC~1977, RCW~120 and RCW~49. We use the model on \cii\ line observations of these regions. Apart from being one of the major coolants of the ISM, the 1.9~THz fine-structure line of ionized carbon, C$^+$ (\cii\,), is also among the brightest lines in Photo Dissociation Regions (PDRs) \citep{Crawford1985,Stacey1991,Bennet1994}. 
PDRs are exposed to FUV photons (6~eV $<$ h$\nu <$ 13.6~eV) where H$_2$ is dissociated to H (hydrogen) and C (carbon) is ionised to C$^+$. Owing to the lower ionisation potential of C than that of H,  \cii\ traces the transition from H$^+$ to H and H$_2$ \citep{Hollenbach1999,Wolfire2022}. Recent observational studies have proved \cii\ to be a powerful diagnostic in improving our understanding of stellar feedback and massive star formation. It probes dynamic structures like expanding shells, photo-radiated surfaces of molecular clouds, and pillars in the ISM \citep{Pabst2019,Pabst2020,Luisi2021,Tiwari2021,Bonne2022,Kavak2022}. Moreover, \citet{Schneider2023} recently reported the role of \cii\ in unveiling the dynamic interactions between cloud ensembles in the Cygnus region of our Galaxy. This makes \cii\ a favourable candidate for this study. \\


In this paper, we provide a step-by-step guide for the community to be able to use the GMM to assist in identifying coherent physical structures (expanding shell/bubbles, interacting molecular clouds, etc) in the ISM. We also discuss the model results on three Galactic sources and evaluate the accuracy of the models in identifying different clusters in a region. We note that throughout this paper, we use the term `clusters' to refer to groups of similar spectra, as described below, and not to clusters of stars.


\begin{figure*}
\centering
\includegraphics[width=180mm]{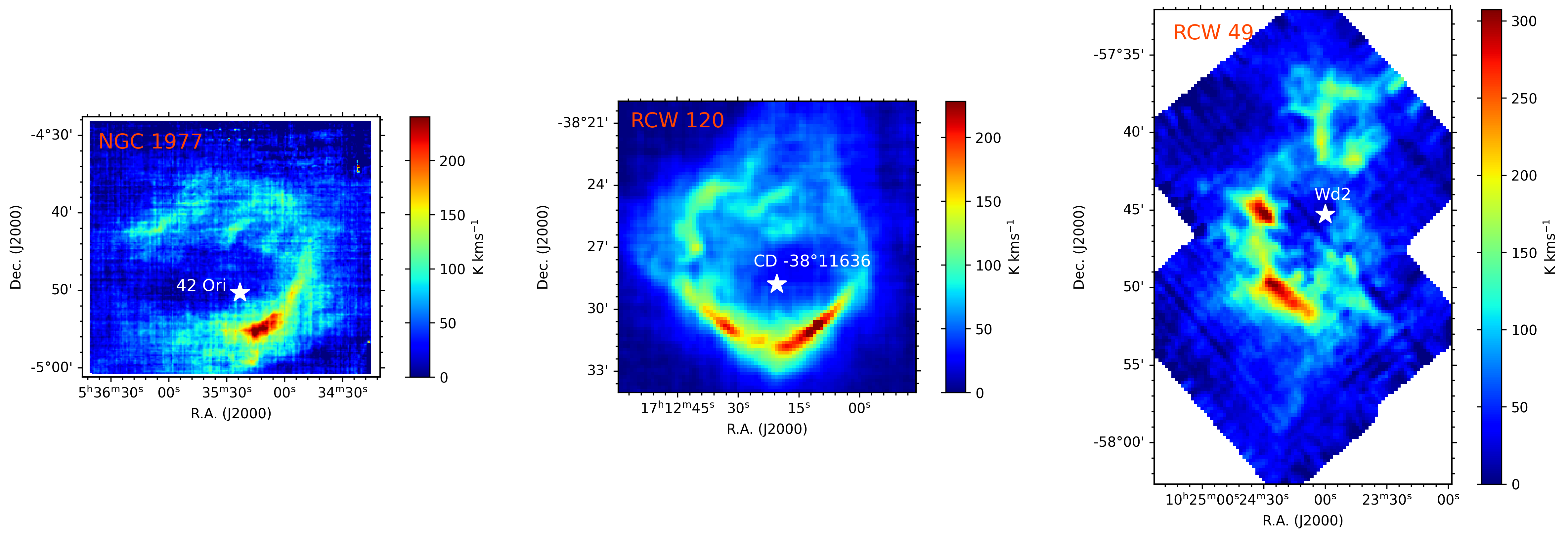}

\caption{Velocity integrated (within the ranges given in Table~\ref{tab:input-parameters}) intensity maps of \cii\ towards NGC~1977, RCW~120 and RCW~49. The pink colored stars mark the main ionizing sources in these regions. These plots are made using regrided data cubes for the three sources based on S/N requirement described in Sec.~\ref{sec:guide}.
\label{fig:vel-int-maps}}
\end{figure*}

\section{Gaussian Mixture Model} \label{sec:gmm}

In principle, machine learning techniques require learning through a training set. In contrast,  Gaussian Mixture Models (GMM) are a standard statistical regression procedure that is best calculated through maximum likelihood with the EM Algorithm \citep{McLachlan2000, Dempster1977, Bouveyron2019, Fruhwirth2019}. Hence, strictly speaking, Gaussian Mixture Models are not a machine learning technique. Nevertheless, in the literature, Gaussian Mixture Models are commonly referred to as an unsupervised machine learning method \citep{Murphy2013} and, here, we adhere to that choice by the community. GMM aims to describe complex high-dimensional data as a linear combination of multidimensional Gaussian distribution \citep{McLachlan2000}. These models are well studied and have been improved by e.g. \citet{Smyth1999, Blei2004, Bovy2011, Melchior2018}. Their various applications indicate the importance of this tool in astronomy and other fields \citep{Greenspan2006, Jones2019, Riaz2020, Kabanovic2022}.



Our observed dataset consists of N spectra, each of which give the main beam temperature ($T_{\rm MB}$) as a function of velocity ($v_{\rm LSR}$). We can interpret each spectrum as a single point $x_i$ in $\mathcal{D}$ dimensional space where $\mathcal{D}$ is the number of velocity channels. 
The GMM attempts to find clusters of spectra that have similar profiles to each other. It does so by assuming the dataset is generated from a limited number ($K$) of Gaussian clusters each with mean $\mu_{\rm k}$ and covariance, $\Sigma_{\rm k}$. The mean of each Gaussian cluster can be interpreted as its average spectrum. By describing the set of all observed spectra as a weighted linear combination of a limited number of clusters, the probabilistic distribution is determined by maximizing the likelihood, defined as:
\begin{equation}
    \ln{\mathcal{L}}\left(\boldsymbol{x}\, |\, \boldsymbol{\theta}\right) = \sum_{i=1}^N\: \ln\: \sum_{k=1}^K\:\alpha_k\:\mathcal{N}(\,x_i\, |\, \mu_k, \Sigma_k).\label{eq:gmm_likelihood}
\end{equation}

Here $\mathcal{N}(\,x_i\, |\, \mu_k, \Sigma_k)$ represents the multidimensional Gaussian distribution and $\alpha_k$ is the weight attributed to that cluster with $\sum_k \alpha_k = 1$. The likelihood is computed by summing over all $K$ Gaussian components for all $N$ spectra individually. The optimization is performed using the Expectation-Maximization (EM) algorithm \citep{Dempster1977}.  
This method has recently been used by \citet{Kabanovic2022} to recognize several distinct physical structures in velocity resolved observations of Atacama Pathfinder EXperiment (APEX) telescope's $^{12}$CO data towards RCW~120.\\

In this work we aim to expand upon earlier research and implement the more advanced GMMis \citep{Melchior2018}. Unlike a traditional GMM this implementation is capable of dealing with noise by enforcing a minimum covariance on the clusters, and of automatically determining an optimal number of clusters by setting the weight of a less probable cluster to extremely low ($\sim10^{-10}$) values. These expansions are achieved through adaptations of the likelihood function and the optimization algorithm. This behaviour of the GMMis can be tuned through a set of hyperparameters. These are different from model parameters (such as each $\mu_k$ and $\Sigma_k$) in that they are set \textit{a priori} and influence the resulting model. Examples of hyperparameters here are the number of provided clusters, the minimum covariance regularization and the stopping criterion. We discuss the considerations and choices for hyperparameters further in Section \ref{sec:hyperparameters}.

The complete algorithms and parameter definitions are described in detail in \cite{Melchior2018} and in their \texttt{PyGMMis} code.\footnote{\href{https://github.com/pmelchior/pygmmis}{https://github.com/pmelchior/pygmmis}} The method and code used to get the results presented in this work are publicly available\footnote{\href{http://hdl.handle.net/1903/30423}{http://hdl.handle.net/1903/30423}} at the digital repository of University of Maryland.

\section{SOFIA Observations} \label{sec:data}


In this work, we use the \cii\,~158~$\mu$m (1.9~THz) observations, which were observed using the upGREAT\footnote{German Receiver for Astronomy at Terahertz. (up)GREAT is a development by the MPI f\"ur Radioastronomie and the KOSMA/Universit\"at zu K\"oln, in cooperation with the DLR Institut f\"ur Optische Sensorsysteme.} \citep{Risacher2018} heterodyne receiver on board the Stratospheric Observatory for Infrared Astronomy (SOFIA, \citealt{Young2012}). On-the-fly maps towards NGC~1977, RCW~120 and RCW~49 were observed in February 2017 and June 2019. RCW~120 and RCW~49 were observed as a part of the SOFIA Legacy program, FEEDBACK \citep{Schneider2020}. 
A map size of 15$\arcmin$ $\times$ 15$\arcmin$ was observed for RCW~120 \citep{Luisi2021,Kabanovic2022} and for RCW~49, the map size was 25.1$\arcmin$ $\times$ 25.1$\arcmin$ \citep{Tiwari2021}. 
For NGC~1977, we cropped out a 36.2$\arcmin$ $\times$ 32.5$\arcmin$ 
region from a 1.15 square degree map of the Orion Nebula complex \citep{Pabst2020} centered on the NGC~1977 bubble.
The native spatial and spectral resolution of the observations at 1.9~THz were 14.1$\arcsec$ and 0.04~km~s$^{-1}$. 
For more observational details see \citet{Schneider2020} and \citet{Higgins2021}.





\section{Sources} \label{sec:sources}
We selected three sources with relatively different complexity in terms of ionising source, morphology and star formation activity. Fig.~\ref{fig:vel-int-maps} shows the velocity integrated intensity maps of NGC~1977, RCW~120 and RCW~49. We introduce the sources below:

\subsection{NGC~1977} \label{sec:ngc1977}

NGC~1977 is the northern-most shell in the Orion Nebula Complex within Orion~A, which is $\sim$ 
395~pc away \citep{Grosschedl2018}. A B1V type star, 42 Orionis, is the main ionising source in this region \citep{Peterson2008}. \cite{Pabst2020} reported a shell in NGC~1977 of radius $\sim$ 1~pc which is expanding at a speed of $\sim$ 1.5~km~s$^{-1}$ and has a mass of $\sim$ 700~\(\textup{M}_\odot\). The thermal pressure of the ionised gas is responsible for powering the expansion of the shell in NGC~1977.
The expansion timescale is estimated to be 0.4~Myr from the \cite{Spitzer1968} solution of thermal expansion of an \hii\ region.


\begin{deluxetable*}{l c c c c c}

\tablecaption{Input and output parameters of the models}
\tablehead{
 \colhead{Sources} &  
 \colhead{rms (or $\omega$)} & 
\colhead{velocity range} & 
 \colhead{$n_{\rm input}$} &
 \colhead{$n_{\rm output}$} &
 \colhead{\textbf{$n_{\rm output\_final}$}}
}
\startdata
NGC~1977  & 0.85 & 5 to 16 & 18 &8 &6\\
RCW~120  & 0.46 & -30 to 10 & 14 &5 &4\\
RCW~49  & 0.29 & -30 to 30 & 16 &10 &5\\
\enddata
\tablecomments{The rms is in the units of K, the velocity range is in the units of km~s$^{-1}$, $n_{\rm input}$ is the number of clusters we give in the models as input, $n_{\rm output}$ is the number of clusters identified by the models and $n_{\rm output\_final}$ is $n_{\rm output}$ excluding noise artefacts and background emission.}
\label{tab:input-parameters}
\end{deluxetable*}

\subsection{RCW~120} \label{sec:rcw120}

RCW~120 is a southern \hii\ region, mainly ionized by a single O8V type star, CD $-$38$\degree$11636 (LSS 3959) \citep{Georgelin1970}, and is located at a distance of 1.7~kpc. 
Two molecular clouds are associated with RCW~120. One is blue-shifted and bright within a velocity range of -37 to -20~km~s$^{-1}$, while the other is red-shifted and bright within a velocity range of -20 to 10~km~s$^{-1}$. 
The near-perfect circular morphology with a prominent opening to the north of RCW~120 as seen in far and mid-infrared wavelengths belongs to the red-shifted cloud component \citep{Zavagno2007,Deharveng2009,Luisi2021,Kabanovic2022}. 
The stellar wind of the O-type stars drives the  expanding shell with a velocity of 15~km~s$^{-1}$ and a mass of up to 520~\(\textup{M}_\odot\) \citep{Luisi2021}. The fast expansion speed of the bubble constrains the lifetime of the region to $\sim$ 0.15~Myr. X-ray observations suggest that hot plasma is venting out to the east and through an opening in the north, such that 20\% of the thermal energy is lost \citep{Luisi2021}.

\subsection{RCW~49} \label{sec:rcw49}
RCW~49 is one of the most luminous star-forming regions in the southern Galaxy and is located at a distance of 4.16~kpc  \citep{Alvarez2013,Zeidler2015,Tiwari2021}. A compact stellar cluster, Westerlund~2 (Wd2), comprises 37 OB stars and $\sim$ 30 early-type OB star candidates around it. Moreover, a Wolf-Rayet (WR) binary (WR20a) is part of the central Wd2 cluster, while another WR star (WR20b) and an O5V star are a few arcminutes away from the cluster center \citep{Ascenso2007,Tsujimoto2007,Rauw2011,Mohr-Smith2015,Zeidler2015}. The stellar winds of the Wd2 cluster and WR20a have swept up a shell of radius 6~pc which expands at a speed of $\sim$ 13~km~s$^{-1}$ and has a mass of 2.5 $\times$ 10$^4$~\(\textup{M}_\odot\). Besides the shell, this region also hosts two large scale molecular clouds whose collision led to the formation of the Wd2 cluster \citep{Furukawa2009}, namely: the ridge and the northern \& southern clouds \citep{Tiwari2021,Tiwari2022}.  

\section{Step-by-step guide to use the Gaussian Mixture Models}\label{sec:guide}

For a given data cube, the Gaussian Mixture Model identifies various physical structures associated with unique spectral profiles depending on the different input parameters we give. The use of different combinations of input parameters can lead to significantly different results. Below, we provide a recipe that users can follow when applying these models to the data on their sources.  

In short, we have a data cube with three dimensions (two spatial, one spectral). We fit a Gaussian Mixture Model with $K$ clusters, decided based on the first breaking point in the $n_{\rm input}$ versus $n_{\rm output}$ diagram. Each spatial pixel is then assigned to one of these clusters based on its probability of being generated by that cluster.

\begin{figure*}
\centering
\includegraphics[width=58mm]{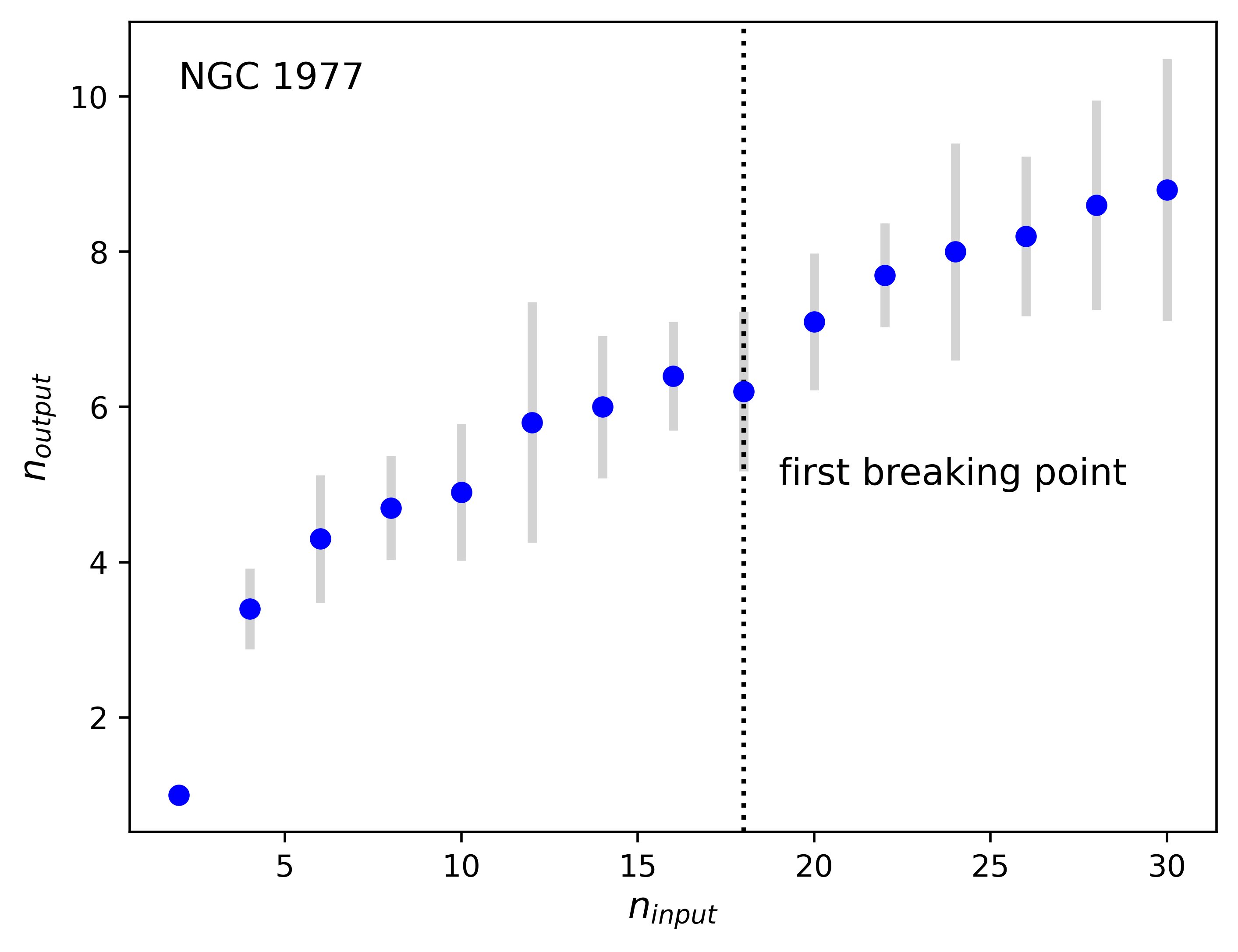}
\includegraphics[width=58mm]{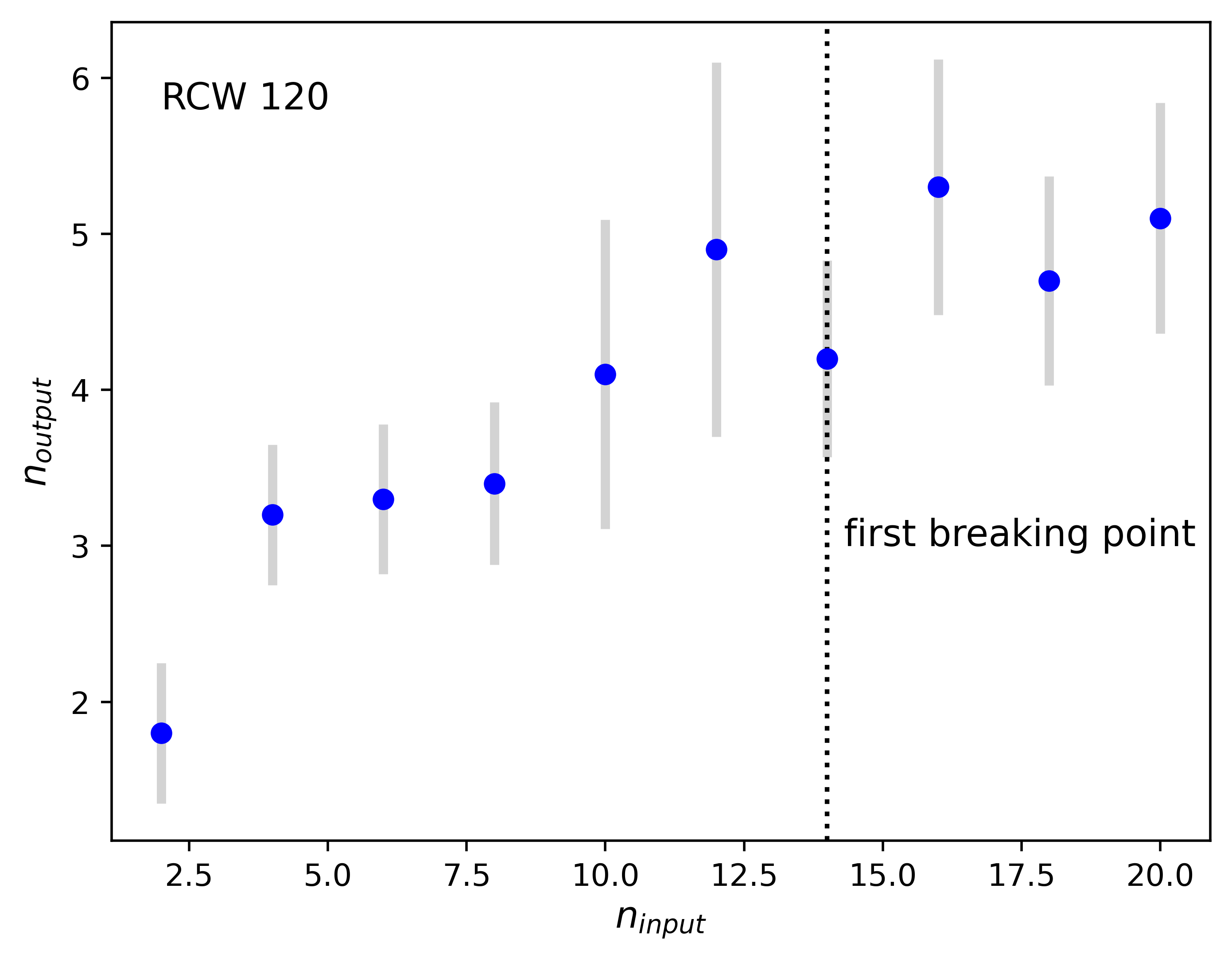}
\includegraphics[width=58mm]{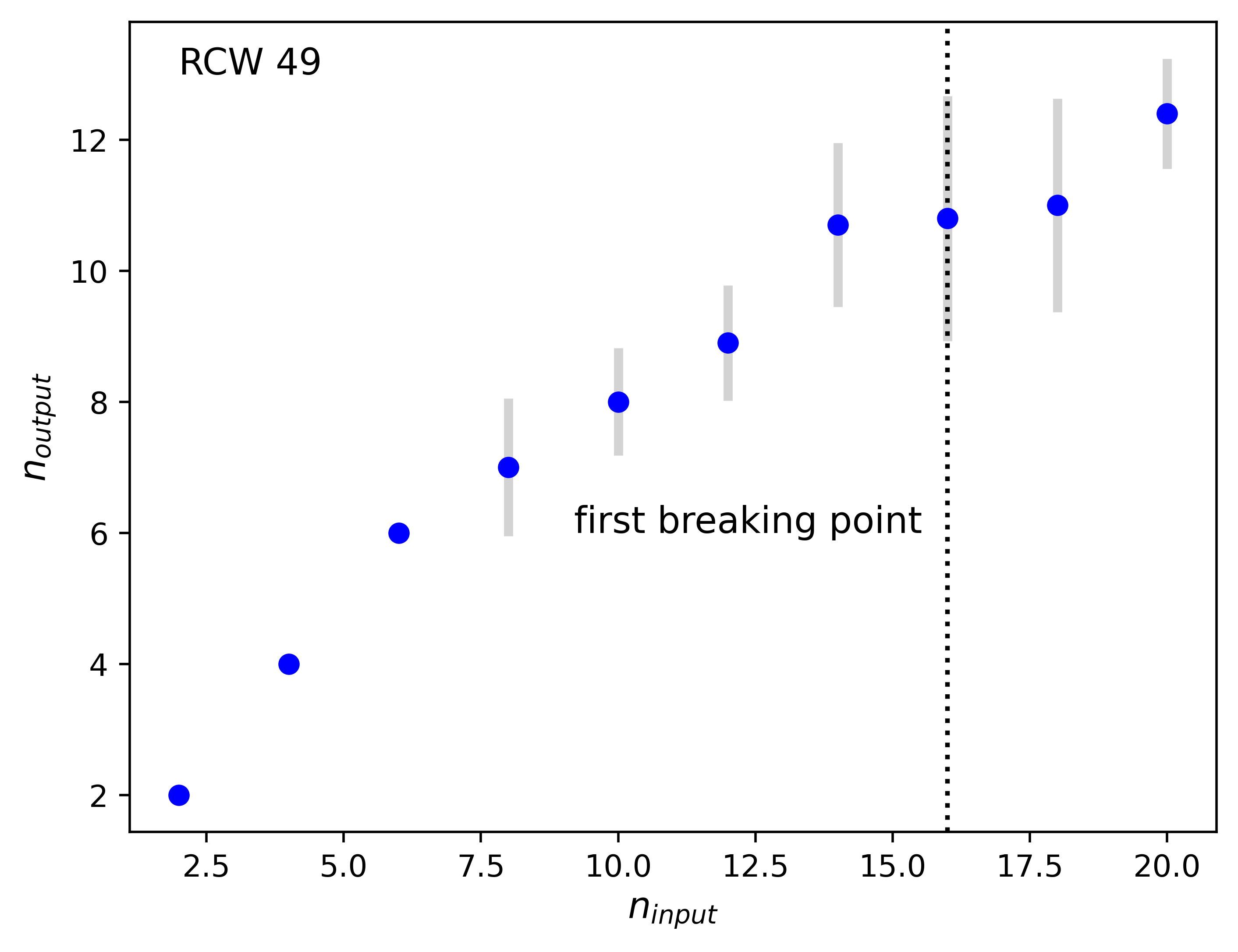}

\caption{Number of clusters identified by GMM ($n_{\rm output}$) versus number of clusters given as input ($n_{\rm input}$) for NGC~1977, RCW~120 and RCW~49. Models were run 10, 50 and 10 times for each $n_{\rm input}$ in NGC~1977, RCW~120 and RCW~49, respectively. Their mean $n_{\rm output}$ is shown with blue, while the standard deviation is displayed in gray.   
\label{fig:n_output-n_input}}
\end{figure*}

\subsection{Signal-to-noise ratio of the data cube}\label{sec:s/n}

It is preferable that the data cube on which the Gaussian Mixture Model is run has a signal-to-noise (S/N) $\gtrsim$ 10. This ensures that the model does not identify noisy pixels as `real' and independent clusters. The noisy pixels mentioned here do not only correspond to the observed map edges that are under-sampled, but rather are spread out all over the map. They seem to have high intensity (or brightness) but are actually artefacts.


The S/N of a cube is determined by:
\begin{equation}
    \frac{S}{N} = \frac{I_{\rm avg}}{rms}, 
\end{equation}

where $I_{\rm avg}$ (in K) is the average intensity of the \cii\ emission in the entire cube area and the rms (in K) of the cube is calculated in an emission free velocity window following the method given in \citet{Higgins2021}.

For cases where the S/N of a cube is $<$ 10, we suggest the users to smoothen the cube to a lower spatial resolution. This can be achieved by gradually increasing the beam size and pixel size following the Nyquist theorem. The cube can also be spectrally regrided to a lower velocity resolution upto a limit of $\frac{Full~width~half~maxima}{3}$. Moreover, if it is not feasible to improve the S/N of a given data set, follow the process described in Sec.~\ref{sec:hyperparameters} point~\ref{sec:cov}.

For NGC~1977, we keep the cube at the spatial resolution of 18$\arcsec$ (following the data reduction technique given in \citealt{Higgins2021}) 
with a velocity resolution of 0.5~km~s$^{-1}$. We find the $I_{\rm avg}$ $\sim$ 12~K and an rms = 0.85~K such that the S/N $\sim$ 14. 

For RCW~120, we regrided the cube to a 20$\arcsec$ spatial resolution with a velocity resolution of 0.5~km~s$^{-1}$. We find the $I_{\rm avg}$ $\sim$ 5~K and an rms = 0.46~K such that the S/N $>$ 10.  

For RCW~49, we regrided the cube to a 25$\arcsec$ spatial resolution with a velocity resolution of 1~km~s$^{-1}$. We find the $I_{\rm avg}$ $\sim$ 3~K and an rms = 0.29~K such that the S/N $\sim$ 10.

\subsection{Hyperparameter variations}\label{sec:hyperparameters}
\begin{enumerate}

\item Velocity range: The observed \cii\ data cubes of 
NGC~1977, RCW~120 and RCW~49 have hundreds of velocity channels i.e. the observations were taken for large velocity windows. However the \cii\ emission itself is confined to smaller velocity ranges i.e. to a smaller number of velocity channels, e.g. NGC~1977 has 619 channels in total but the \cii\ emission is constrained to 50 channels. Therefore, we constrain the algorithm
to only fit its models on these channels or velocity range. The velocity ranges for the three sources are given in Table~\ref{tab:input-parameters}.

\item rms threshold: Any velocity channel which has an rms evaluated over the entire source area lower than this limit is discarded before we fit the model. We essentially throw out any channels that contain little information to decrease computing time. A too low value can cause the model to be unable to converge within a reasonable time, and a too high value might result in real clusters disappearing. After extensive experimentation we set this value to 3 $\times$ rms for all sources presented in this work.

\item Normalization: Using the Gaussian Mixture Models, we want to identify the main coherent physical structures in a source. Depending on the scales of the data, machine learning techniques might require normalisation of the dataset to increase model performance.
We have tested the models using several normalisation techniques, but it is important to note that the choice of normalization technique can bias the resulting model. When implementing such a method, its impact needs to be carefully considered based on both the data and the desired result. \citet{Kabanovic2022} opted for \textit{min-max} normalization which places a larger emphasis on the spectral shape over the peak intensities. We have investigated the results after applying \textit{mean} normalization (see Appendix~\ref{app:mean}) but opt for no normalization to limit the amount of bias we introduce.






\item Covariance regularisation, $\omega$\label{sec:cov}: 
This is essentially a spectral intensity limit. $\omega$ describes a minimum $T_{\rm mb}$ below which the model is not allowed to explain variances in the data. This technique is implemented by \citet{Melchior2018} based on the work by \citet{Bovy2011} and equips GMM with the power to suppress the effects of noise. The traditional GMM (used in e.g. \citealt{Kabanovic2022}) does not include $\omega$ as an hyperparameter.  
We use $\omega$ = rms where S/N $\gtrsim$ 10. In cases where the S/N $<$ 10 for a source and the data quality cannot be enhanced through various reduction techniques, we suggest to use $\omega$ equal to the standard deviation of the data points of an rms map of a source. 

\item Number of clusters: 
When running any clustering algorithm, a number of clusters needs to be provided \textit{a priori}. The traditional GMM (Appendix~\ref{app:slawa-models}), as used by e.g. \citet{Kabanovic2022}, must use all provided clusters ($n_{\rm input} = n_{\rm output}$), and an information criterion is required to determine the optimal $n_{\rm input}$. Examples include the BIC and AIC \citep[Bayesian and Akaike Information Criterion;][]{Schwarz1978} which penalize the likelihood of a model by the number of clusters it uses. This is not required in the GMMis method used here, since an adjustment in the Expectation-Maximization algorithm allows arbitrarily low cluster weights \citep[See][ for mathematical details]{Bovy2011, Melchior2018}. This means that $n_{\rm output} \leq n_{\rm input}$. Nevertheless, an optimal $n_{\rm input}$ still needs to be determined. Too few will lead to loss of information, while too many can lead to assignments of insignificant clusters and increasingly long computing times. Therefore, we propose a new method of determining this optimal value.

The value $n_{\rm output}$ generally increases as $n_{\rm input}$ increases, i.e. the models identify more coherent physical structures in a source. However, this trend breaks for certain $n_{\rm input}$ and we define the first occurrence of this as the `first breaking point'. 
This can be seen in Fig.~\ref{fig:n_output-n_input}, where $n_{\rm output}$ versus $n_{\rm input}$ is plotted for all three sources. Based on the quantitative increase in $n_{\rm output}$ when increasing $n_{\rm input}$ linearly, we incremented $n_{\rm input}$ as a multiple of 2 for all sources. 
Quantitatively, a breaking point occurs at the minimum of the first derivative of $n_{\rm output}$ versus $n_{\rm input}$ i.e. minima of $\Delta n_{\rm output}/\Delta n_{\rm input}$. 
Models were run multiple times for every $n_{\rm input}$, and the standard deviations of the resulting $n_{\rm output}$ are shown as error bars (in gray) in Fig.~\ref{fig:n_output-n_input}.
For NGC~1977, RCW~120 and RCW~49, the first breaking point occurs at $n_{\rm input}$ = 18, 14 and 16 clusters, respectively. 


\end{enumerate}

\begin{figure}
    \centering
    \includegraphics[width=80mm]{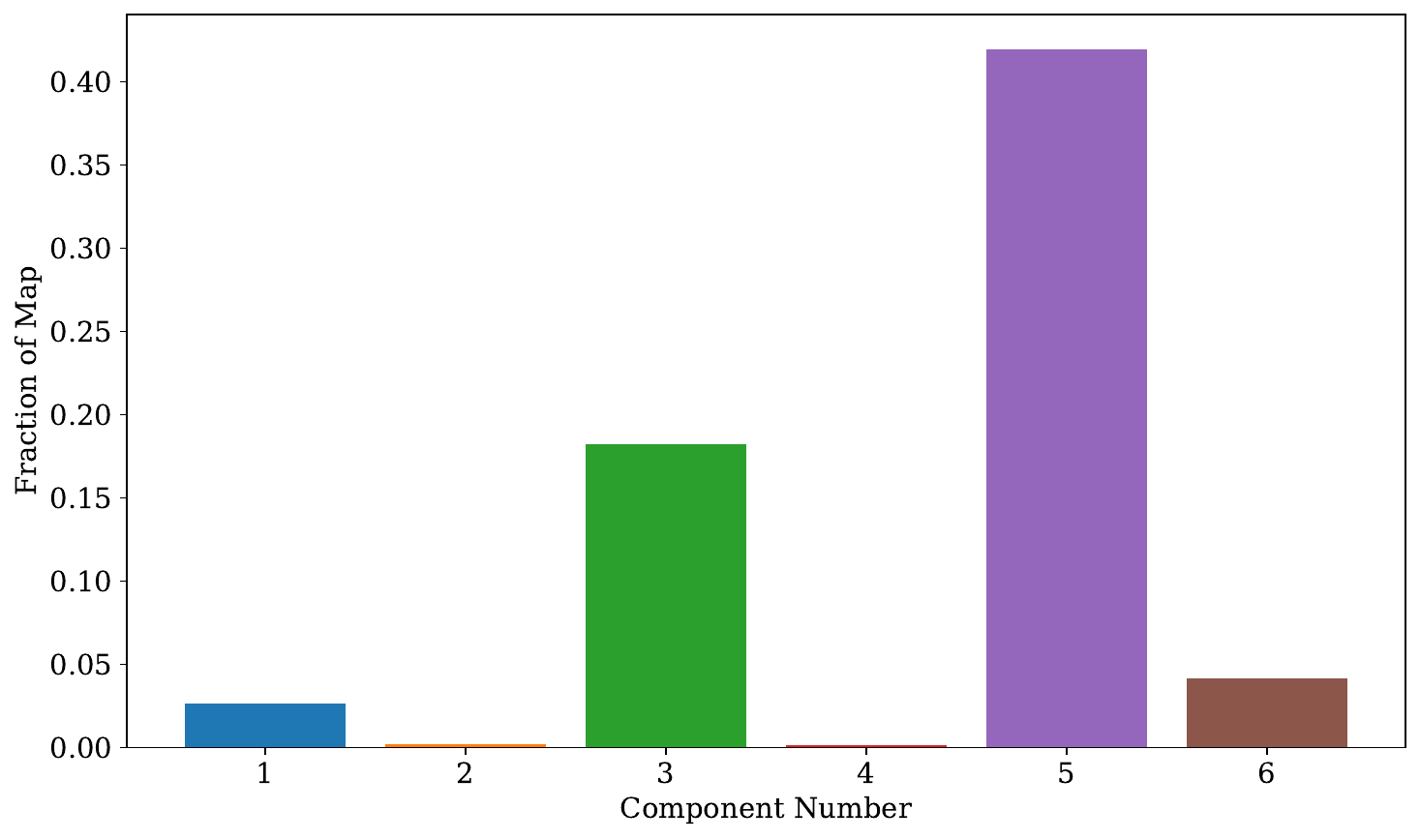}
    \includegraphics[width=80mm]{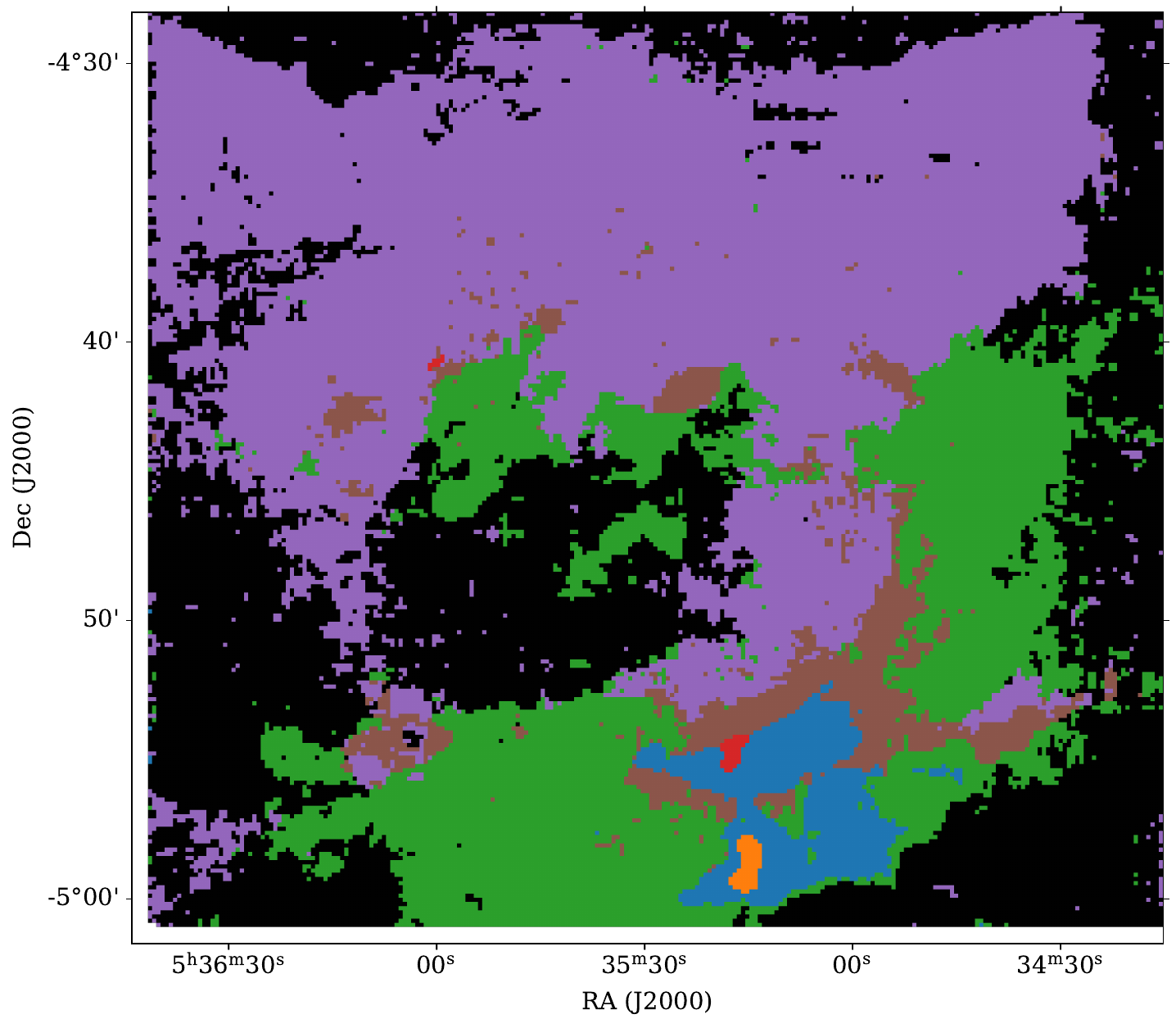}
    
    \caption{GMM results of NGC~1977: weights map (top) and domain map (bottom) displaying the identified clusters.}
    \label{fig:ngc1977-wieights-dm-spec}
\end{figure}

\begin{deluxetable*}{c c c c c c c}
    \tablecaption{Computed physical quantities for GMM clusters of NGC~1997}
    \tablehead{
    \colhead{cluster} &  
    \colhead{size (pc$^2$)} & 
    \colhead{velocity component} &
    \colhead{flux (K~km~s$^{-1}$)} & 
    \colhead{position (km~s$^{-1}$)} &
    \colhead{width (km~s$^{-1}$}) &
    \colhead{T$_{\rm peak}$ (K)}
    }
    \startdata
    1 & 0.44 & 1 & 124.00 (0.24) & 11.04 (0.00) & 3.64 (0.01) & 32.05 \\
    2 & 0.02 & 1 & 135.85 (0.77) & 10.27 (0.01) & 4.01 (0.03) & 31.81 \\
    3 & 3.08 & 1 & 63.72 (0.21) & 11.39 (0.01) & 3.15 (0.01) & 19.01 \\
    4 & 0.01 & 1 & 166.37 (0.98) & 11.75 (0.01) & 3.46 (0.02) & 45.13 \\
    5 & 6.40 & 1 & 39.68 (0.17) & 12.61 (0.01) & 2.59 (0.01) & 14.40 \\
    6 & 0.74 & 1 & 94.00 (0.20) & 11.80 (0.01) & 3.42 (0.01) & 25.81 \\
    \enddata
    \tablecomments{Columns are from left to right the cluster number, size of the cluster, velocity component of the average spectra, velocity integrated intensity, centroid LSR velocity, full width half maxima of the spectra, peak temperature. The error bars for the flux, position and width are given in brackets.}

    \label{tab:ngc1977-quantities}
\end{deluxetable*}

\section{Gaussian Mixture Model results}\label{sec:GMM-results}
Based on the input parameters (set following the discussion in Sect.~\ref{sec:hyperparameters}
and given in Table~\ref{tab:input-parameters}), the models identify a fixed number of spectral profiles of \cii\ emission for a source. Every pixel of a data cube has 0 to 1 probability to be assigned to one of these spectral profiles. We choose to assign it to the cluster $k$ where this probability is largest. For any individual pixel the probabilities over each cluster $k \in K$ sum up to 1.

\begin{figure}
\centering
\includegraphics[width=80mm]{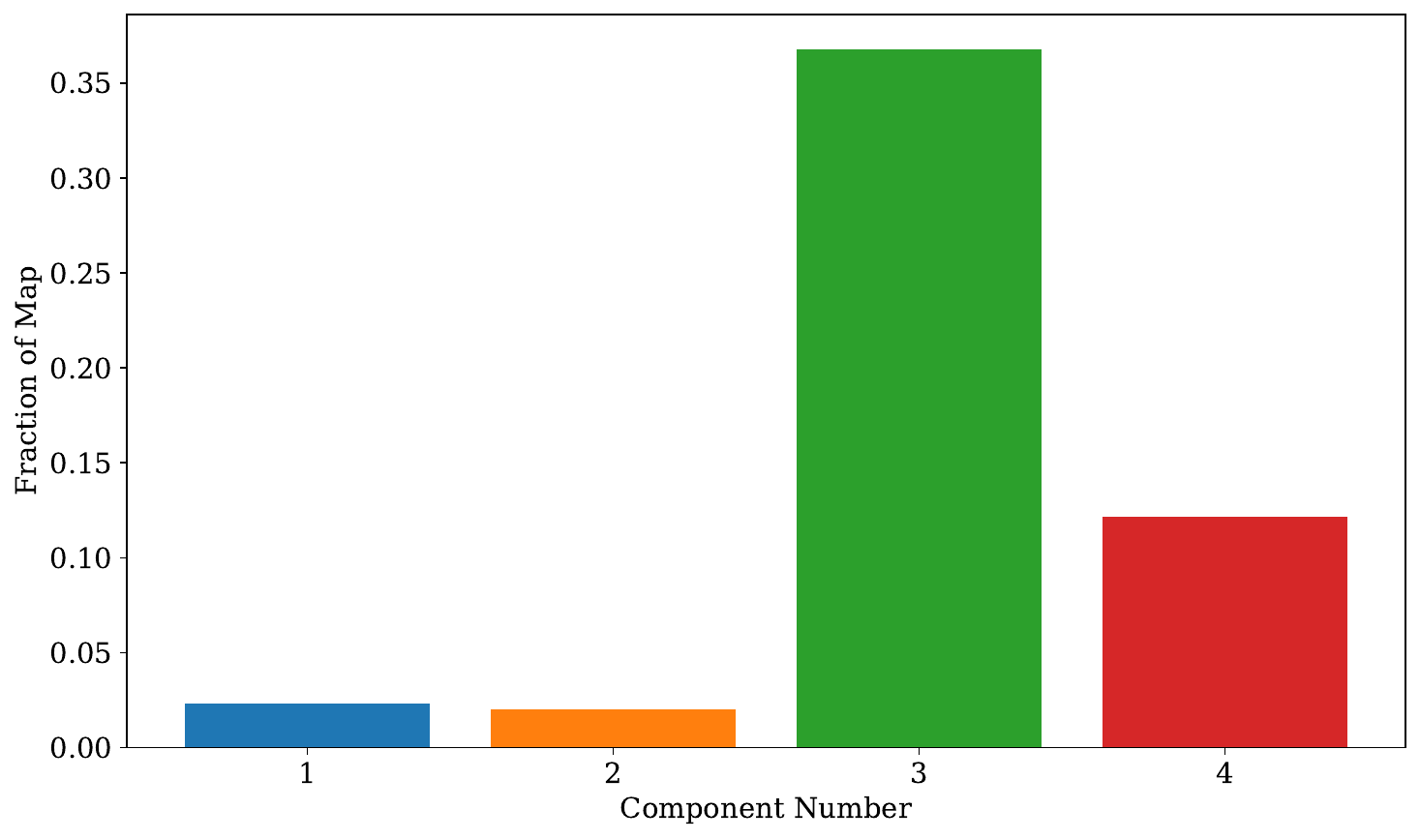}
\includegraphics[width=80mm]{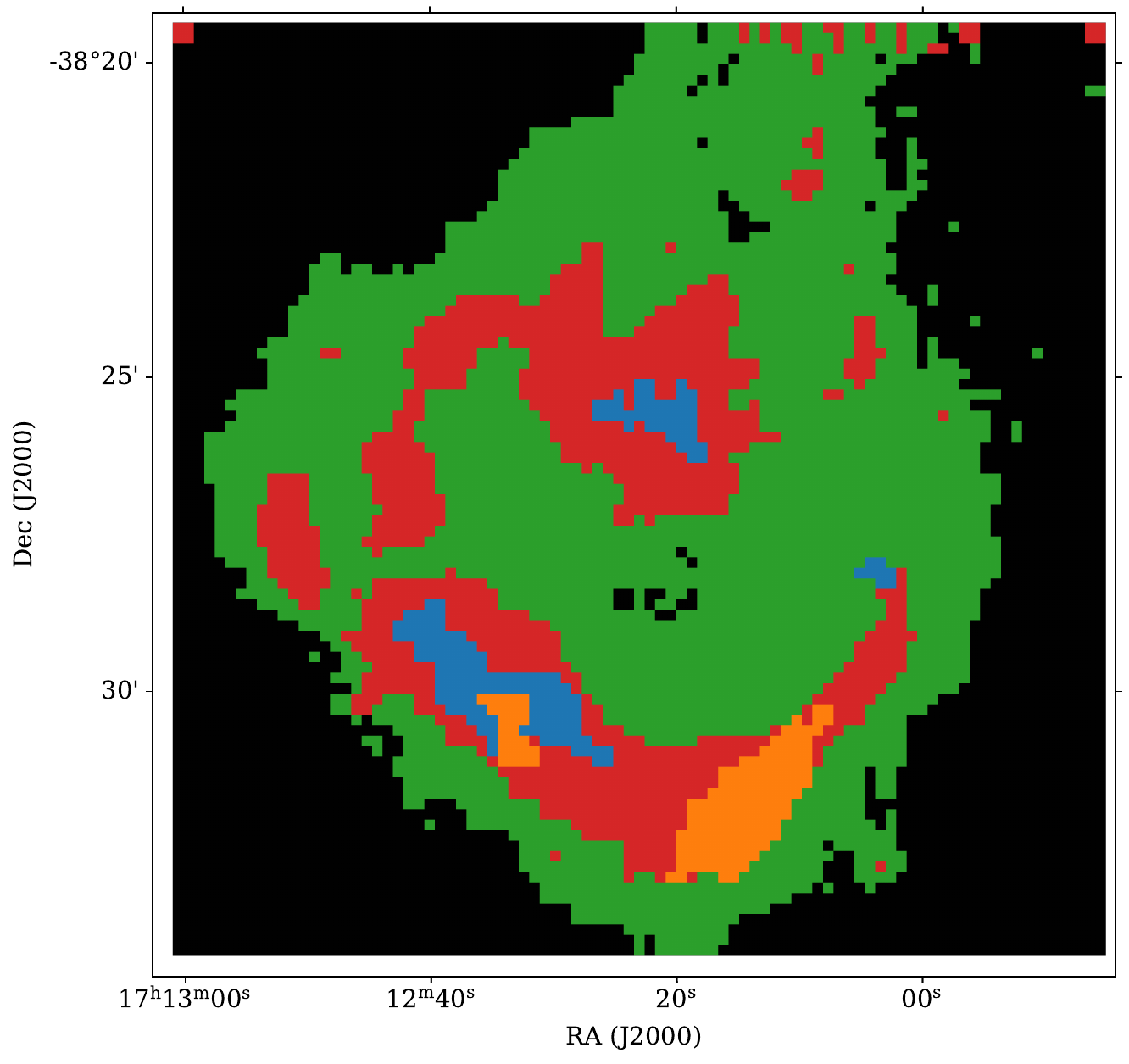}

\caption{GMM results of RCW~120: weights map (top) and domain map (bottom) displaying the identified clusters. 
\label{fig:rcw120-weights-dm-spec}}
\end{figure}

\begin{figure}
\centering
\includegraphics[width=75mm]{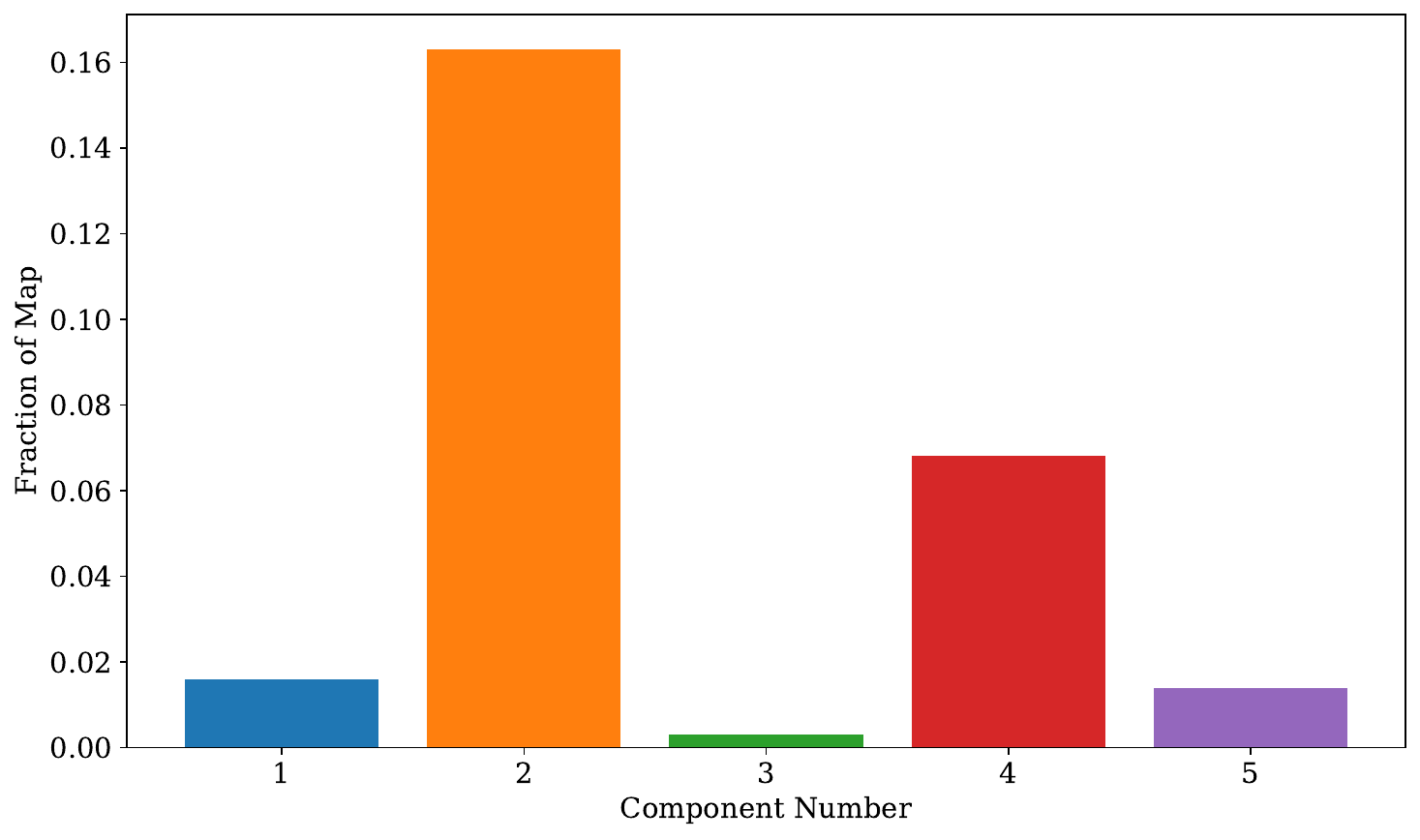}
\includegraphics[width=75mm]{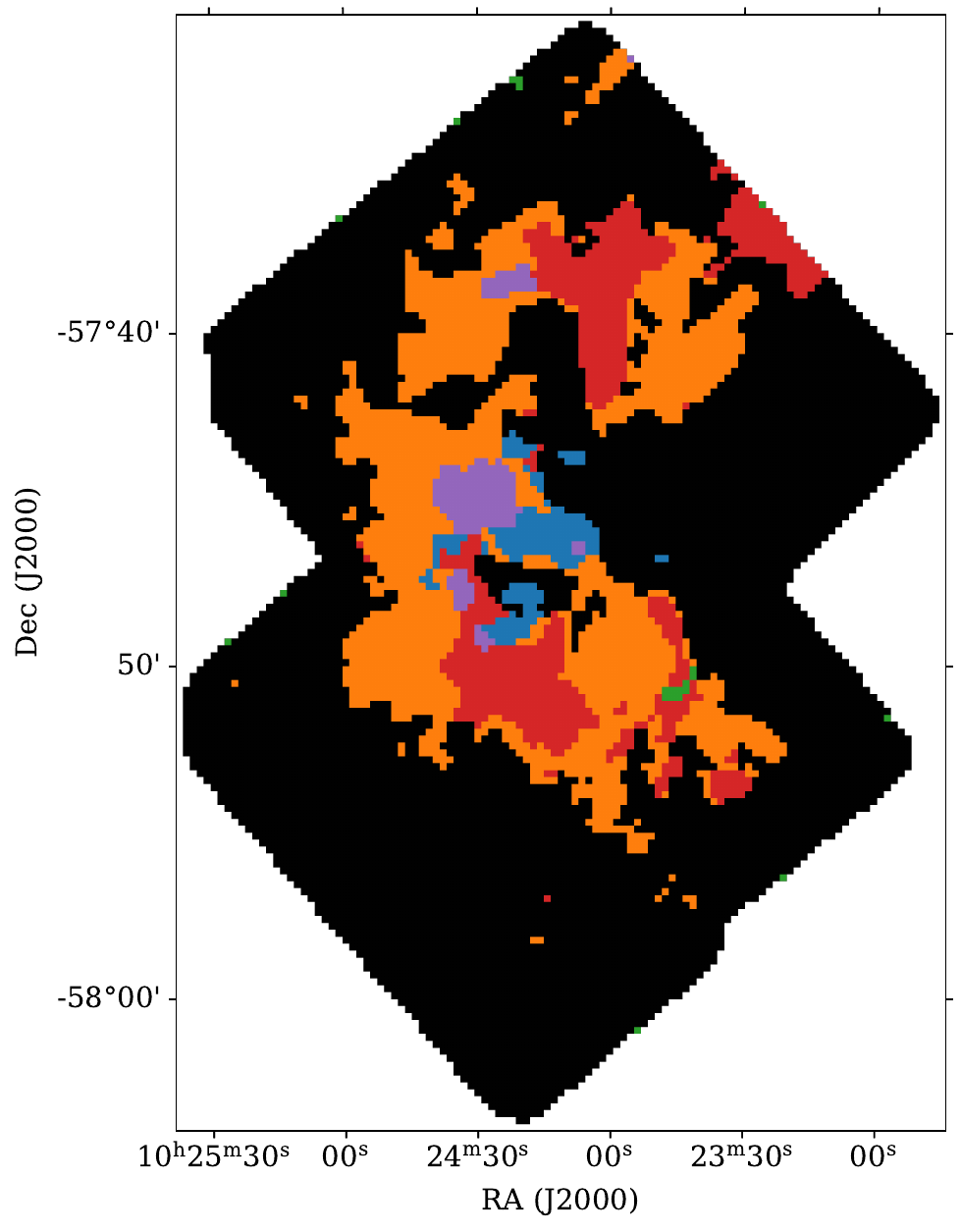}

\caption{GMM results of RCW~49: weights map (top) and domain map (bottom) displaying the identified clusters.
\label{fig:rcw49-weights-dm-spec}}
\end{figure}

\begin{figure}
\centering
\includegraphics[width=85mm]{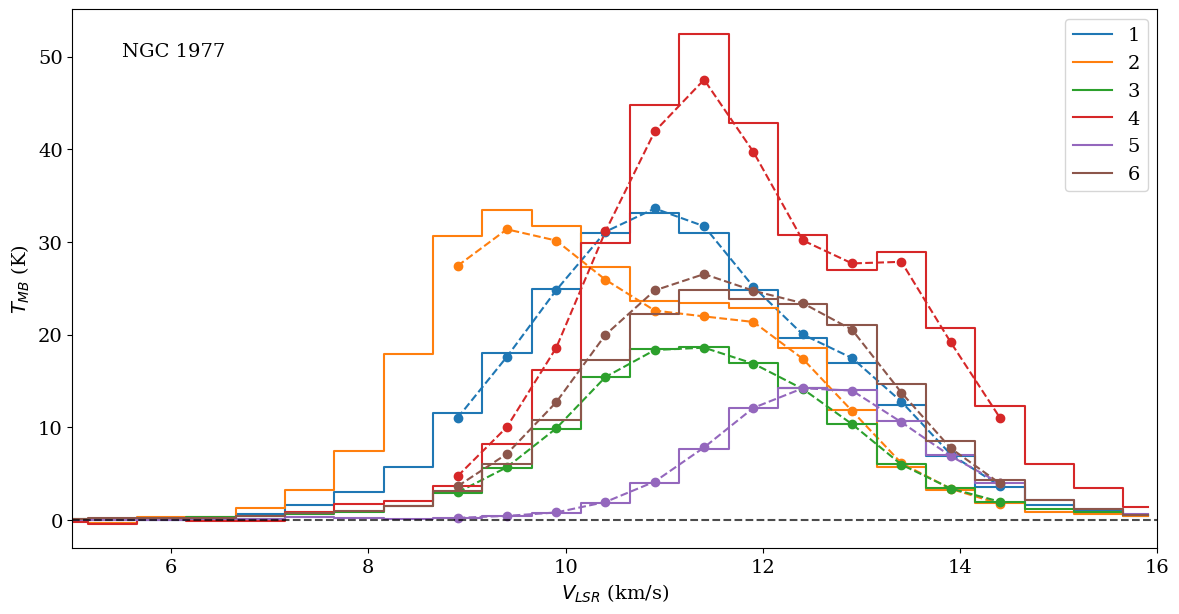}
\includegraphics[width=85mm]{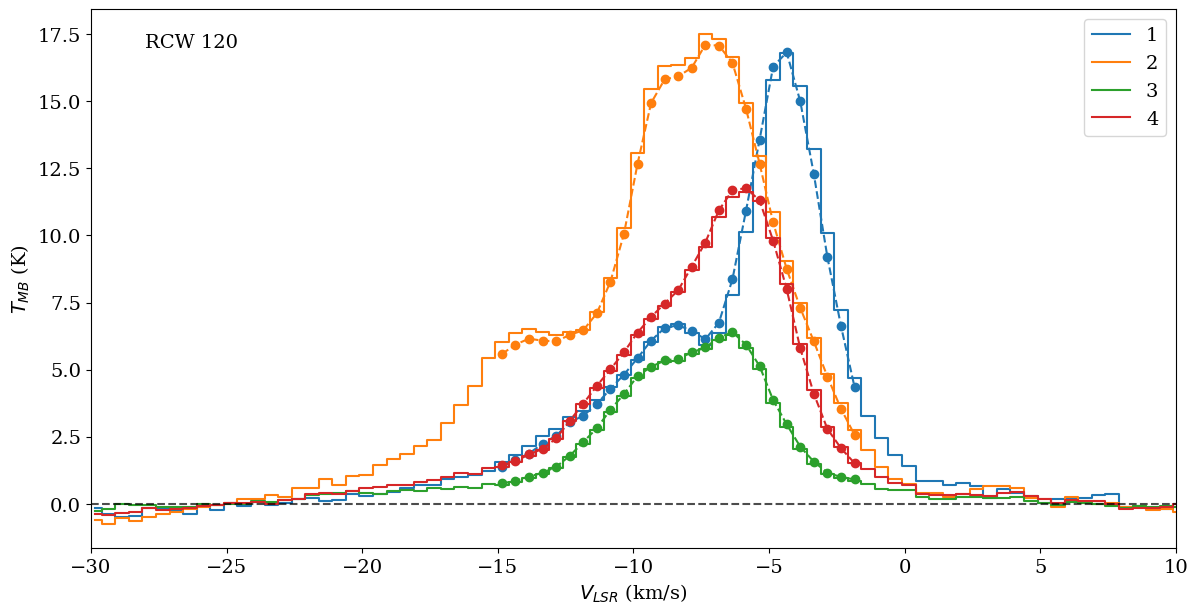} 
\includegraphics[width=85mm]{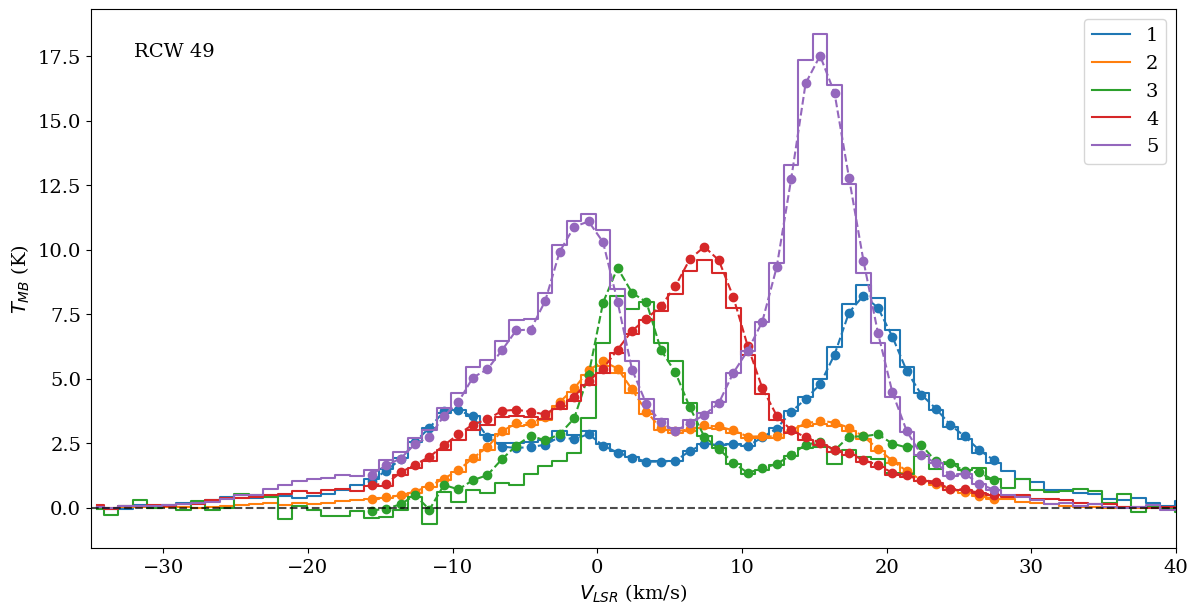}
\caption{Average spectra from all clusters shown in the domain maps of NGC~1977 (top), RCW~120 (middle) and RCW~49 (bottom). The color of every cluster shown here is the same as shown in Figs.~\ref{fig:ngc1977-wieights-dm-spec}, \ref{fig:rcw120-weights-dm-spec} and \ref{fig:rcw49-weights-dm-spec}. The dotted lines indicate the mean of each GMM cluster ($\mu_k$ in eq. \ref{eq:gmm_likelihood}). Channels without dots are those discarded by the RMS threshold. Large differences between the GMM means and the averaged spectra can be an indication of large variance within a single cluster. 
\label{fig:spec}}
\end{figure}

Using these cluster assignments we created weights maps, which show $\alpha_k$ for each cluster. This is analogous to the fraction of data points assigned to cluster $k$. We also made domain maps which show the spatial distributions of the clusters. These results are presented in Figures~\ref{fig:ngc1977-wieights-dm-spec}, \ref{fig:rcw120-weights-dm-spec} and \ref{fig:rcw49-weights-dm-spec}. Tables~\ref{tab:ngc1977-quantities}, \ref{tab:rcw120-quantities} and \ref{tab:rcw49-quantities} list the estimated physical quantities of the identified clusters. The size was determined by counting the number of pixels per cluster and then converting it to a physical scale (kpc) using the distance to NGC~1977, RCW~49 and RCW~120. The flux $\int T_{\rm peak} dv$, position, width ($dv$) and peak temperature ($T_{\rm peak}$) are derived by fitting Gaussians to the average spectra of every cluster (Fig.~\ref{fig:spec}). In cases where more than one velocity components are seen, multiple Gaussians were fitted.\\




For NGC~1977 (Fig.~\ref{fig:ngc1977-wieights-dm-spec}), 8 clusters were identified in total, but we excluded two clusters (black space in the domain map) as one of these traced striped noise artefacts in the top right corner of the cube, and the other contained background emission. 
The extended shell described by \citet{Pabst2020} is traced by the purple and green clusters in Fig.~\ref{fig:ngc1977-wieights-dm-spec}.
The northern and southern halves of the shell are assigned to these two separate clusters because the models identified a velocity gradient throughout the shell which red-shifts the northern half by a few km~s$^{-1}$.
The shell and its velocity gradient can be seen in the \cii\ velocity channel maps shown and discussed in Fig.~\ref{fig:chan-maps-ngc1977} and Appendix~\ref{app:vel-chan-maps}, respectively.

The region towards the molecular core, OMC3, and the PDR created by interaction with the integral-shaped filament \citep{YoungOwl2002} 
is mostly outlined by one cluster (blue).
The two emission peaks are each assigned to their own separate cluster (red and orange), and these clusters have distinctly different velocity structures.
Additionally, through the absence of clusters, we can clearly see that the bubble is very thin on the Eastern side.\\ 


\begin{deluxetable*}{c c c c c c c}
    \tablecaption{Computed physical quantities for GMM clusters of RCW~120.}
        \tablehead{
    \colhead{cluster} &  
    \colhead{size (pc$^2$)} & 
    \colhead{velocity component} &
    \colhead{flux (K~km~s$^{-1}$)} & 
    \colhead{position (km~s$^{-1}$)} &
    \colhead{width (km~s$^{-1}$}) &
    \colhead{T$_{\rm peak}$ (K)}
    }  
    \startdata
     \multirow{2}{*}{1}  &  \multirow{2}{*}{1.07} & 1 & 70.57 (0.79) & - 7.38 (0.06) & 10.72 (0.11) & 6.18 \\
      &      & 2 & 34.82 (0.54) & - 4.17 (0.01) &  2.75 (0.03) & 11.89 \\
    \multirow{2}{*}{2} & \multirow{2}{*}{0.98} & 1 & 29.99 (1.21)  & -14.87 (0.08) &  5.41 (0.24) & 5.21 \\
      &      & 2 & 122.46 (1.17)  & - 7.42 (0.03) &  6.50 (0.07) & 17.53 \\
    3 & 21.13& 1 & 45.65 (0.51) & - 7.68 (0.04) &  7.10 (0.09) & 6.04 \\ 
    4 & 6.97 & 1 & 81.21 (0.74) & - 6.92 (0.03) &  7.22 (0.08) & 10.57 \\
    \enddata
    \tablecomments{Columns are from left to right the cluster number, size of the cluster, velocity component of the average spectra, velocity integrated intensity, centroid LSR velocity, full width half maxima of the spectra, peak temperature. The error bars for the flux, position and width are given in brackets.}
    \label{tab:rcw120-quantities}
\end{deluxetable*}

For RCW~120 (Fig.~\ref{fig:rcw120-weights-dm-spec}), 5 clusters were identified in total, but we excluded one cluster which only described noise and was mainly localised at the edges of the map. 
The circular structure (green colored cluster) consists of the expanding shell and the outer ring emission of RCW~120. The ring dynamics can also be seen in the \cii\ velocity channel maps shown Fig.~\ref{fig:chan-maps-rcw120}.  
\citet{Luisi2021} reported that the shell is broken open in the north and in the east. 
The green cluster in the domain map has an extension in the north-west corresponding to this break in the shell. 
The red colored cluster mainly traces the dense, ring-shaped PDR/torus of RCW~120, from which most of the \cii\ emission originates.
Within the boundary of this red cluster is the expanding shell of RCW~120. In the domain map (Fig.~\ref{fig:rcw120-weights-dm-spec}), the expanding shell is not clearly separated since the bulk emission (green cluster) dominates the spectral shape. However, the expanding shell is somewhat better separated/localized in Fig.~\ref{fig:rcw120-slawa} using the traditional GMM, which suggest an additional cluster.
We find an opening towards the east where the PDR is disrupted. This aligns well with the X-ray emission from RCW~120 (Fig.~3 of \citealt{Luisi2021}). Moreover, \citet{Kabanovic2022} refers to this feature as a `narrow opening' in the east.
Local density enhancements along the PDR/torus are identified as additional clusters by the model.
Thus, we expect these clusters to have different physical conditions. This is also validated by the significant differences in the observed intensities towards different clusters. The orange and blue clusters have higher intensities compared to the other clusters, as seen in the observed average spectrum within each cluster in the middle panel of Fig.~\ref{fig:spec}.\\


\begin{deluxetable*}{c c c c c c c}
    \tablecaption{Computed physical quantities for GMM clusters of RCW~49.}
    \tablehead{
    \colhead{cluster} &  
    \colhead{size (pc$^2$)} & 
    \colhead{velocity component} &
    \colhead{flux (K~km~s$^{-1}$)} & 
    \colhead{position (km~s$^{-1}$)} &
    \colhead{width (km~s$^{-1}$}) &
    \colhead{T$_{\rm peak}$ (K)}
    }
    \startdata
  \multirow{4}{*}{1} & \multirow{4}{*}{11.76} & 1 & 34.48 (4.70) & - 9.85 (0.52) &  9.33 (1.31) &  3.47 \\
      &      & 2 & 19.28 (5.50) & - 0.70 (0.56) &  6.79 (1.58) &  2.67 \\
      &      & 3 & 14.01 (3.40) &   7.35 (0.63) &  6.24 (1.25) &  2.11 \\
      &      & 4 & 84.17 (2.56) &  18.73 (0.13) & 10.52 (0.44) &  7.52 \\
    \multirow{4}{*}{2} & \multirow{4}{*}{121.76} & 1 & 19.55 (0.63) & - 6.68 (0.02) &  7.83 (0.33) &  2.35 \\
      &      & 2 & 38.70 (0.83) &   0.61 (0.07) &  7.15 (0.19) &  5.08 \\
      &      & 3 & 8.85 (0.77) &   7.69 (0.15) &  4.59 (0.29) &  1.81 \\
      &      & 4 & 39.58 (0.60) &  15.30 (0.12) & 11.69 (0.14) &  3.18 \\
    \multirow{2}{*}{3} & \multirow{2}{*}{1.33} & 1 & 44.32 (1.75) &   2.57 (0.06) &  5.94 (0.20) &  7.01 \\
      &      & 2 & 61.03 (3.08) &  14.21 (0.81) & 29.29 (1.31) &  1.96 \\
    \multirow{3}{*}{4} & \multirow{3}{*}{52.11} & 1 & 42.56 (1.03) & - 6.27 (0.09) &  12.92 (0.39) &  3.09 \\
      &      & 2 & 98.73 (1.09) &   6.53 (0.06) & 10.72 (0.08) &  8.65 \\
      &      & 3 & 13.18 (1.13) &  19.79 (0.36) &  9.30 (0.87) &  1.33 \\
     \multirow{2}{*}{5} &  \multirow{2}{*}{10.30} & 1 & 141.72 (2.95) & - 2.30 (0.04) & 13.91 (0.38) &  9.57 \\
     &       & 2 & 128.11 (1.98) &  15.32 (0.05) &  7.12 (0.14) & 16.89 \\
    \enddata
    \tablecomments{Columns are from left to right the cluster number, size of the cluster, velocity component of the average spectra, velocity integrated intensity, centroid LSR velocity, full width half maxima of the spectra, peak temperature. The error bars for the flux, position and width are given in brackets.
    }
    \label{tab:rcw49-quantities}
\end{deluxetable*}

For RCW~49 (Fig.~\ref{fig:rcw49-weights-dm-spec}), 10 clusters were identified in total. Similar to RCW~120, we excluded five that corresponded to noise and were mainly localised on the edges of the map. 
Observational studies (\citealt{Furukawa2009,Tiwari2021,Tiwari2022}) identified 4 different physical structures in RCW~49. We identify three of these.  
The domain map (bottom panel of Fig.~\ref{fig:rcw49-weights-dm-spec}) shows the shell of RCW~49 in orange, which is broken open in the west (reported in \citealt{Tiwari2021}).
It also identifies the `ridge', which corresponds to the purple and blue colored clusters. 
The `northern and southern clouds' reported in Fig.~7 of \citep{Tiwari2021} are identified as the red colored cluster.
The `pillar', which is reported as an independent structure in \citet{Tiwari2022}, is identified as a part of the southern cloud here (Fig.~\ref{fig:rcw49-weights-dm-spec}). In previous studies (\citealt{Tiwari2021, Tiwari2022}), the pillar is reported to be most intense in the velocity range of 5 to 15~km~s$^{-1}$, which overlaps with the southern cloud (most intense within the velocity range of 2 to 8~km~s$^{-1}$). Morphologically, this pillar can be recognized as a separate structure, but kinematically, it is part of the southern cloud.
Despite the rather complex nature of RCW~49, GMM recognizes all relevant structures identified in the in-depth study by \citet{Tiwari2022}. These structures in RCW~49 can also be seen in the \cii\ velocity channel maps shown in Fig.~\ref{fig:chan-maps-rcw49} and their dynamics are described in Appendix~\ref{app:vel-chan-maps}.
The complexity in the velocity structure of RCW~49 can be seen in the \cii\ spectra, where the entire emission spans from -30 to 30~km~s$^{-1}$ (bottom panel, Fig.~\ref{fig:spec}) compared to RCW~120 (-20 to 5~km~s$^{-1}$, middle panel, Fig.~\ref{fig:spec}) and NGC~1977 (6 to 16~km~s$^{-1}$, top panel, Fig.~\ref{fig:spec}). Not only is the velocity range of the emission larger but the number of different velocity components within this emission is largest in RCW~49.

\section{Evaluating the models}\label{sec:model-evaluation} 
In the previous sections, we 
described the steps required to use the models and reported the model results for three relatively different sources. The next natural task is to examine the performance of these models. For this, it is critical to analyze the observed spectra of these regions. Fig.~\ref{fig:spec} shows the observed average spectra of all identified clusters in NGC~1977, RCW~120 and RCW~49 overlaid with dashed curves corresponding to the mean $\mu_k$ of each cluster in each velocity channel. 
Additionally, the dispersion in each cluster-averaged spectrum is shown in Figs.~\ref{fig:ave-spec-std} and \ref{fig:ave-spec-norm-std}.
We find that the dispersion for every spectral profile is small. 
For more details, see Appendix~\ref{app:avg-spec-std}.
 \\

In NGC~1977 the shell is seen in the velocity range 9 to 15~km~s$^{-1}$ \citep{Pabst2020}. The spectra belonging to clusters 3, 5 and 6 (Fig.~\ref{fig:spec}, top panel) fit well into this range. These are the green, purple and brown clusters respectively in the domain map which trace the majority of the limb-brightened shell. 
Spectra 1, 2 and 4 belong to the blue, orange and red clusters towards OMC3 in the domain map. They are visibly brighter and show a double peaked structure which is possibly caused by \cii\ self-absorption (\citealt{Pabst2020} and see \citealt{Guevara2020,Kabanovic2022} for details on optical depth effects).   \\

In RCW~120, the circular ring (bulk emission of RCW~120) is localised within the velocity range of -12 to 2~km~s$^{-1}$ and peaks at -7.5~km~s$^{-1}$ \citep{Kabanovic2022}. The spectra belonging to clusters 3 and 4 (Fig.~\ref{fig:spec}, middle panel) fit this description. They correspond to the green and red clusters, respectively, in the domain map of RCW~120 (Fig.~\ref{fig:rcw120-weights-dm-spec}), which correctly represent the ring. 
Also, the spectrum belonging to cluster 1 (blue colored cluster in the domain map of RCW~120) is similar to the spectra presented by \citet{Kabanovic2022} (their Fig.~5) towards the brightest emission region in \cii\,.\\

In RCW~49, the expanding shell broken open to the west lies within the velocity range of -25 to 0~km~s$^{-1}$ \citep{Tiwari2021}. The average spectrum belonging to cluster 2 (Fig.~\ref{fig:spec}, bottom panel) has its brighter component within this velocity range. It corresponds to the orange cluster in the domain map of RCW~49 (Fig.~\ref{fig:rcw120-weights-dm-spec}), which outlines the broken shell's structure. The northern and southern clouds of RCW~49 are brightest within 2 to 8~km~s$^{-1}$ \citep{Tiwari2021}. This matches with the red colored cluster in the domain map of RCW~49 that corresponds to the average spectrum of cluster 4 in Fig.~\ref{fig:spec}, bottom panel. 
The ridge of RCW~49 has the brightest \cii\ emission within the velocity window of 16 to 22~km~s$^{-1}$ \citep{Tiwari2021}. This matches with the brighter components of the average spectra belonging to clusters 1 and 5. These correspond to the blue and purple clusters in the domain map of RCW~49, which spatially match with the location of the ridge as described in \citep{Tiwari2021,Tiwari2022}.
Furthermore, the average spectrum of cluster 5 in purple has another blue-shifted velocity component,  which corresponds to the shell of RCW~49 and this also agrees well with representative spectrum of position p3 in \citet{Tiwari2022} (and the Fig.~3 within), where both the ridge and shell components overlap.\\

It is important to note that the clusters identified by the GMM do not have strict boundaries. While most of the pixels of an observed \cii\ map corresponding to a given source are assigned to a specific cluster with probability $>$ 0.9 (1 being the highest), at the boundaries of the identified clusters the probabilities can be lower leading to more spurious assignments. Thus, spatial transition from one cluster to another can be more gradual than the strict boundaries illustrated by the domain maps (Figs.~\ref{fig:ngc1977-wieights-dm-spec},\ref{fig:rcw120-weights-dm-spec} and \ref{fig:rcw49-weights-dm-spec}). 
As illustrated in Appendix~\ref{app:low-prob}, the importance of such ill-defined assignments can be assessed by comparing the map of lower probability pixels in a cluster to the map of the cluster. For NGC~1977, this comparison clearly illustrates that pixels with ill-defined cluster status are relegated to the boundary of the cluster and indicate a somewhat more gradual transition from one cluster to the next.

\section{Conclusions}\label{sec:conclusions}
The objective of this study was to identify coherent physical structures in the ISM through an automated technique, which will assist the astronomical community in analysing large scale observations. We investigated the capabilities of the GMM in achieving this goal by testing the models on three different Galactic sources. \\

We ran the GMM on NGC~1977, RCW~120 and RCW~49. From expressing the need of a S/N $>$ 10 for the input data cube to choosing the right number of input clusters into the models, we prepared a step-by-step guide for the users to try GMM on their data sets. The models identified 6, 4 and 5 clusters (excluding noise and background clusters) in NGC~1977, RCW~120 and RCW~49, respectively, and these results were illustrated using weight and domain maps. In NGC~1977, the models identified the expanding shell as well as the dense PDR tracing the interaction of this shell with the OMC-3 core and the Integral-Shaped Filament, consistent with previous work on this source. The domain map also revealed a break in the shell towards the east. In RCW~120, the models identified the bulk emission and the shell, which is broken in the north, and also confirmed another leak in the east, which was identified in the X-ray data earlier. In RCW~49, the models identified the broken shell, the ridge, and the northern \& southern clouds, which agrees with the previous observational studies towards this region. 

We also 
validated the models by examining the average spectrum of each cluster and found that the models assign the data cube pixels to the `right' cluster with high accuracy.  
Successful comparisons between the results of our work and those in literature serve to validate the precision of these models in identifying major physical structures within a given region.
This, in turn, reveals the promise of this powerful and easy-to-use method for analysing and interpreting large \cii\ data sets, such as those to be delivered by NASA's GUSTO and ASTHROS balloon missions.\\

We thank the anonymous referees for bringing important issues to our attention and helping to clarify the paper.

This study was based on observations made with the NASA/DLR
Stratospheric Observatory for Infrared Astronomy (SOFIA). SOFIA is
jointly operated by the Universities Space Research Association
Inc. (USRA), under NASA contract NNA17BF53C, and the Deutsches SOFIA
Institut (DSI), under DLR contract 50OK0901 to the University of
Stuttgart. upGREAT is a development by the MPI f\"ur Radioastronomie and
the KOSMA/University of Cologne, in cooperation with the DLR Institut
f\"ur Optische Sensorsysteme.

N.S. acknowledges support from the FEEDBACK-plus project that is
supported by the BMWI via DLR, Project number 50OR2217 (FEEDBACK-plus).
S.K. acknowledges support from the Orion-Legacy project that is
is supported by the BMWI via DLR, project number 50OR2311.
Publication costs were provided by NASA through award SOF070077 issued by USRA.


\bibliography{references}{}
\bibliographystyle{aasjournal}

\appendix

\begin{figure}[h]
\centering
\includegraphics[width=85mm]{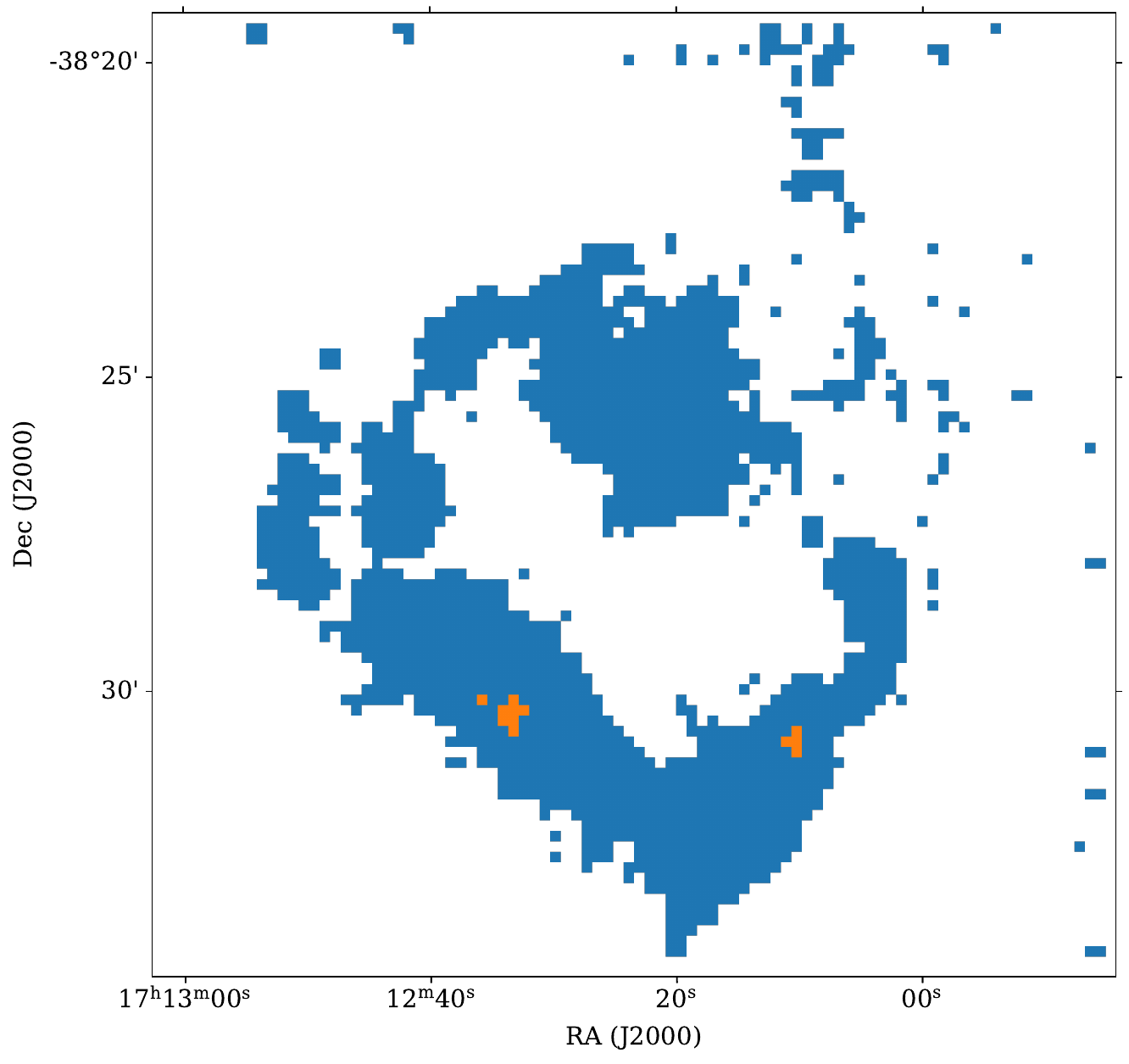}

\caption{Domain map of RCW~120 displaying the identified clusters with $mean$ normalisation technique. The other hyper-parameters were same as those given in Table~\ref{tab:input-parameters}. 
\label{fig:rcw120-dm-mean}}
\end{figure}

\begin{figure}[h]
\centering

\includegraphics[width=50mm]{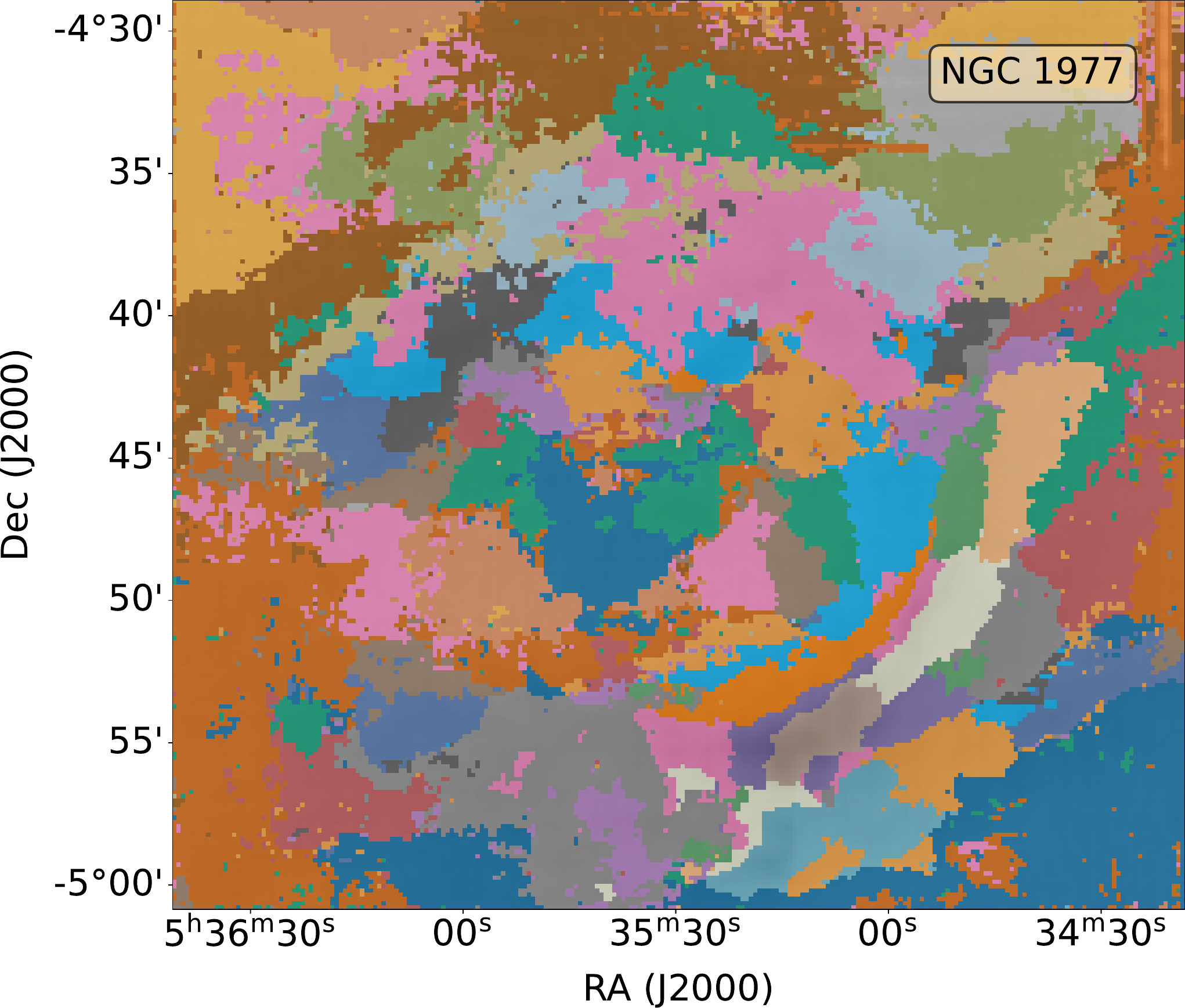}\quad\quad
\includegraphics[width=50mm]{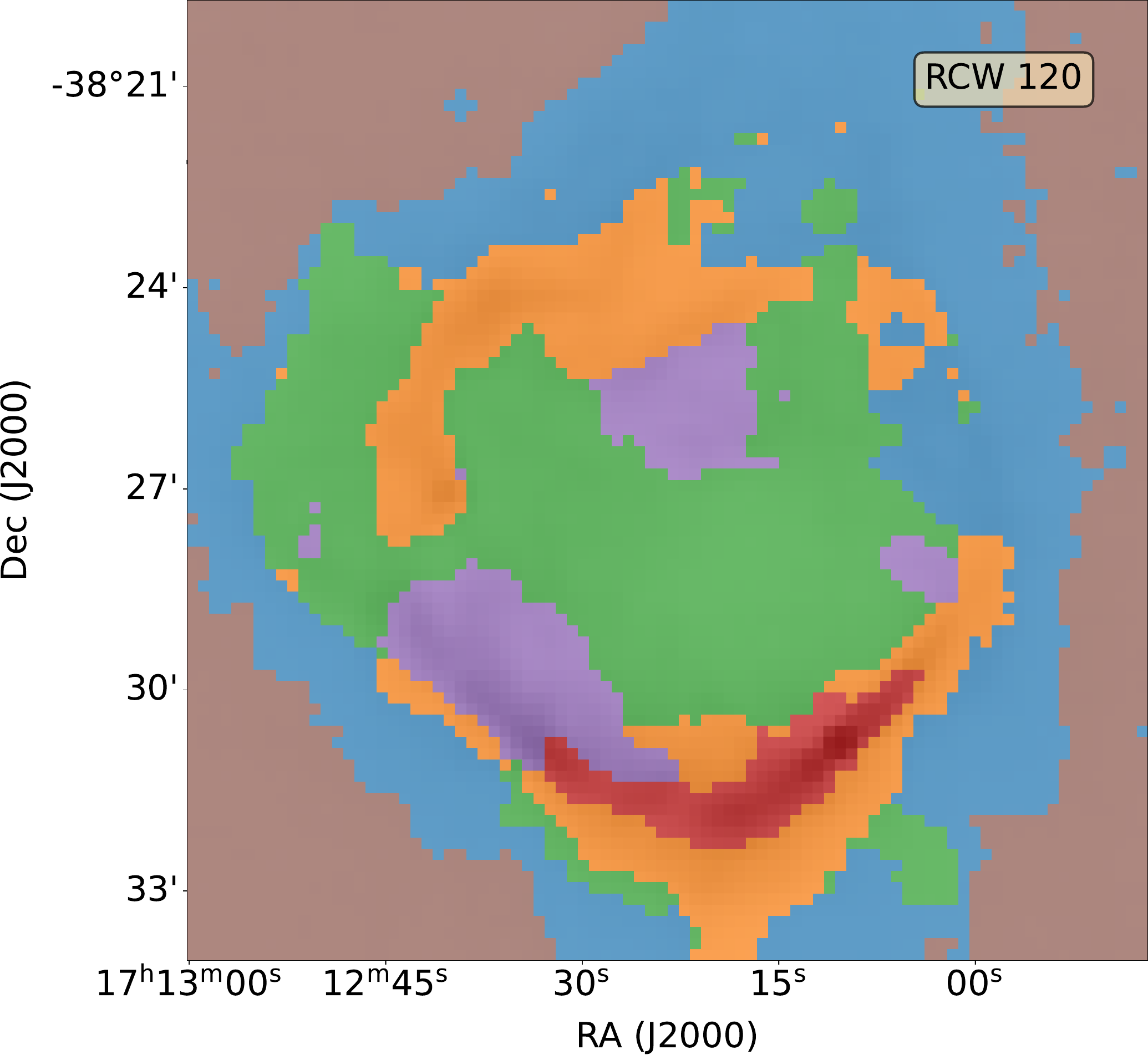}\quad\quad
\includegraphics[width=50mm]{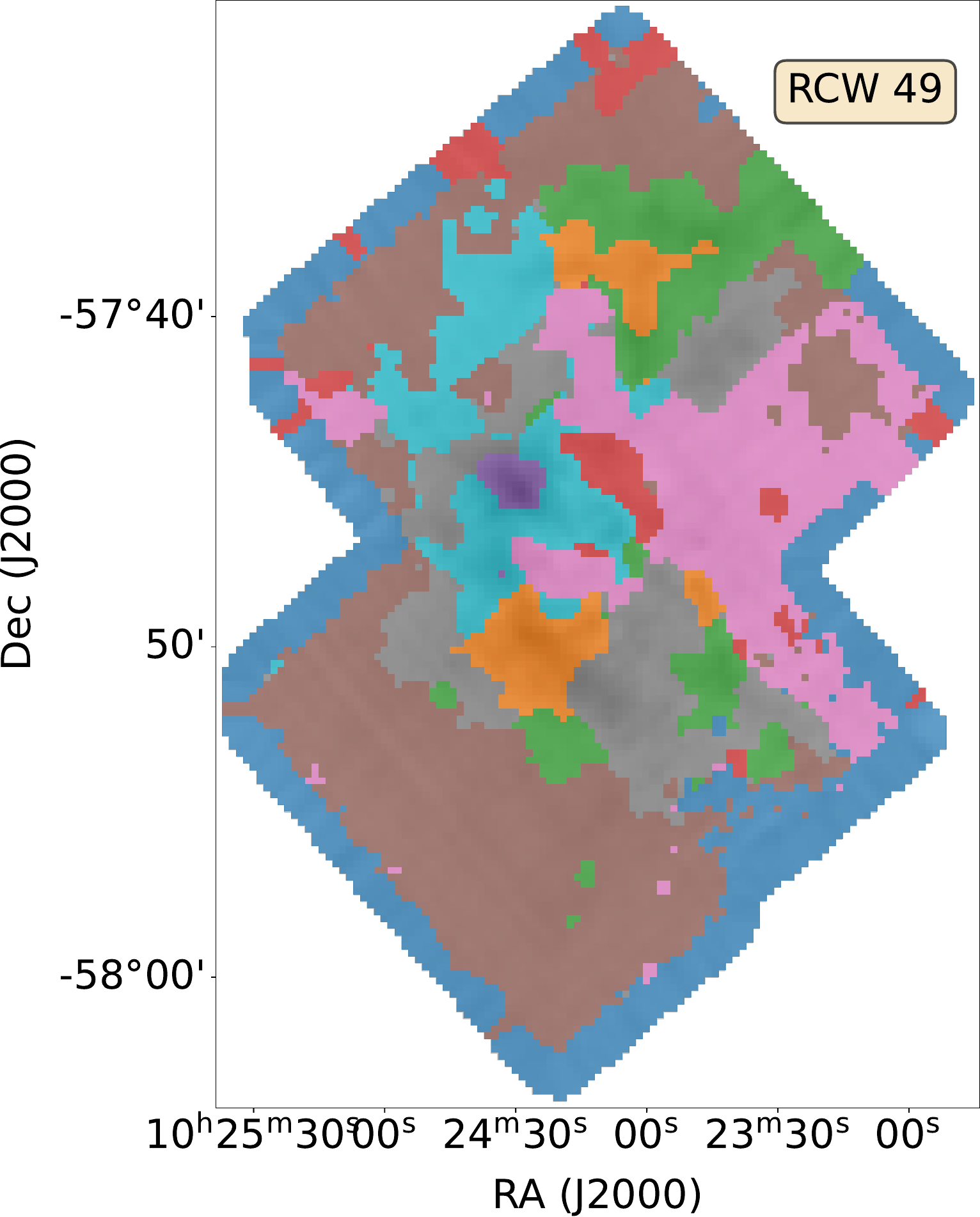}
\includegraphics[width=50mm]{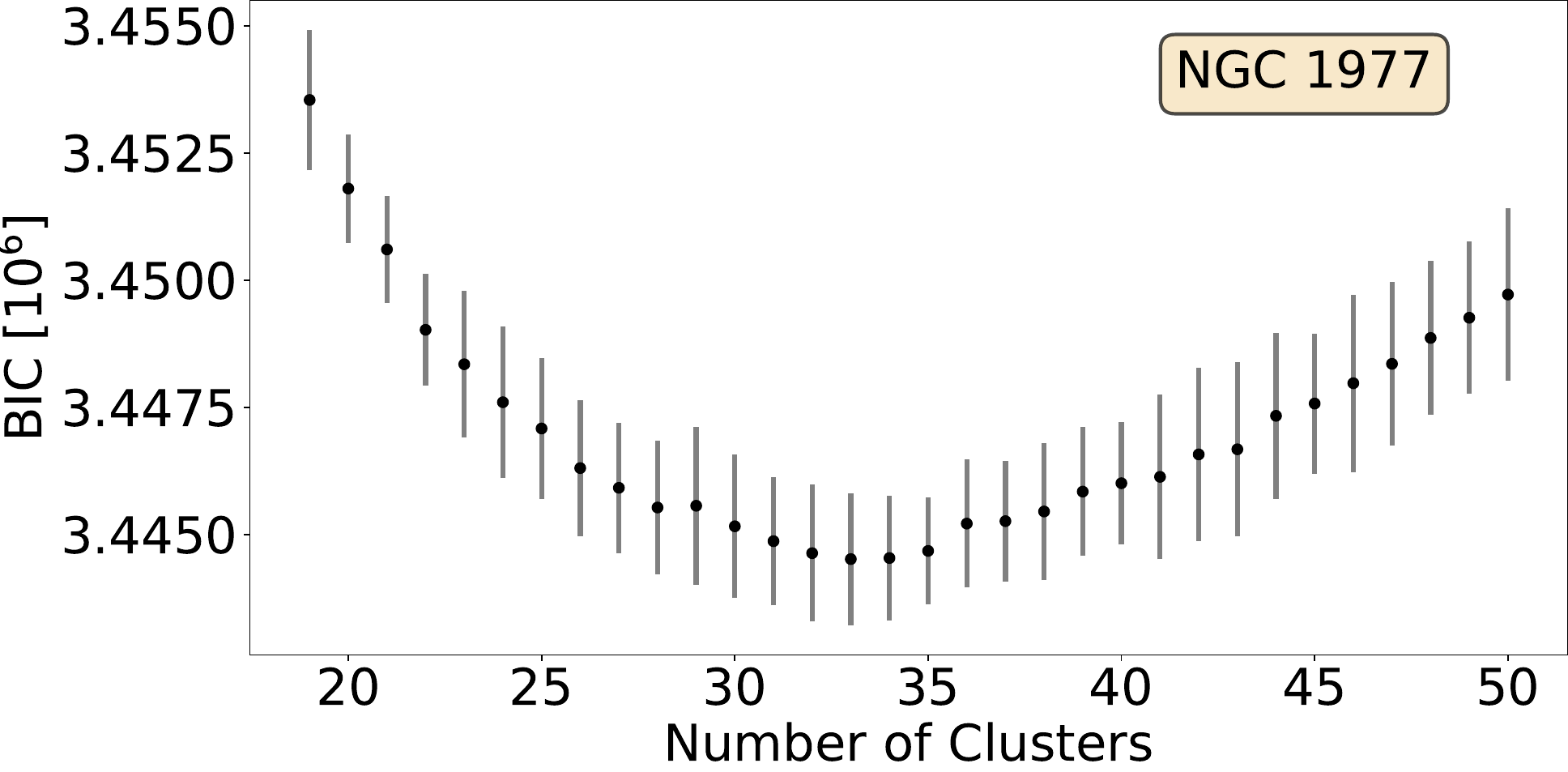}\quad\quad
\includegraphics[width=50mm]{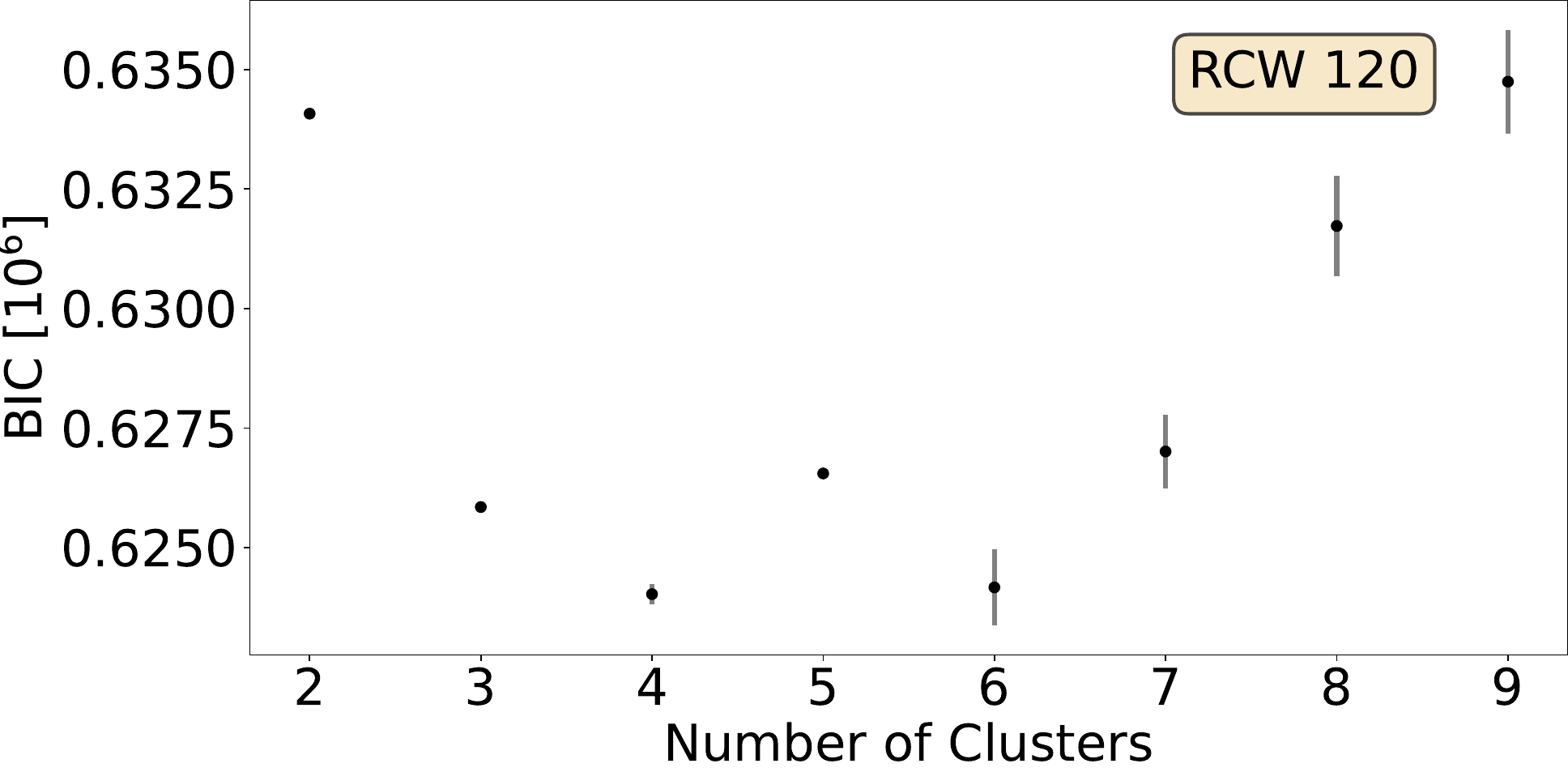}\quad\quad
\includegraphics[width=50mm]{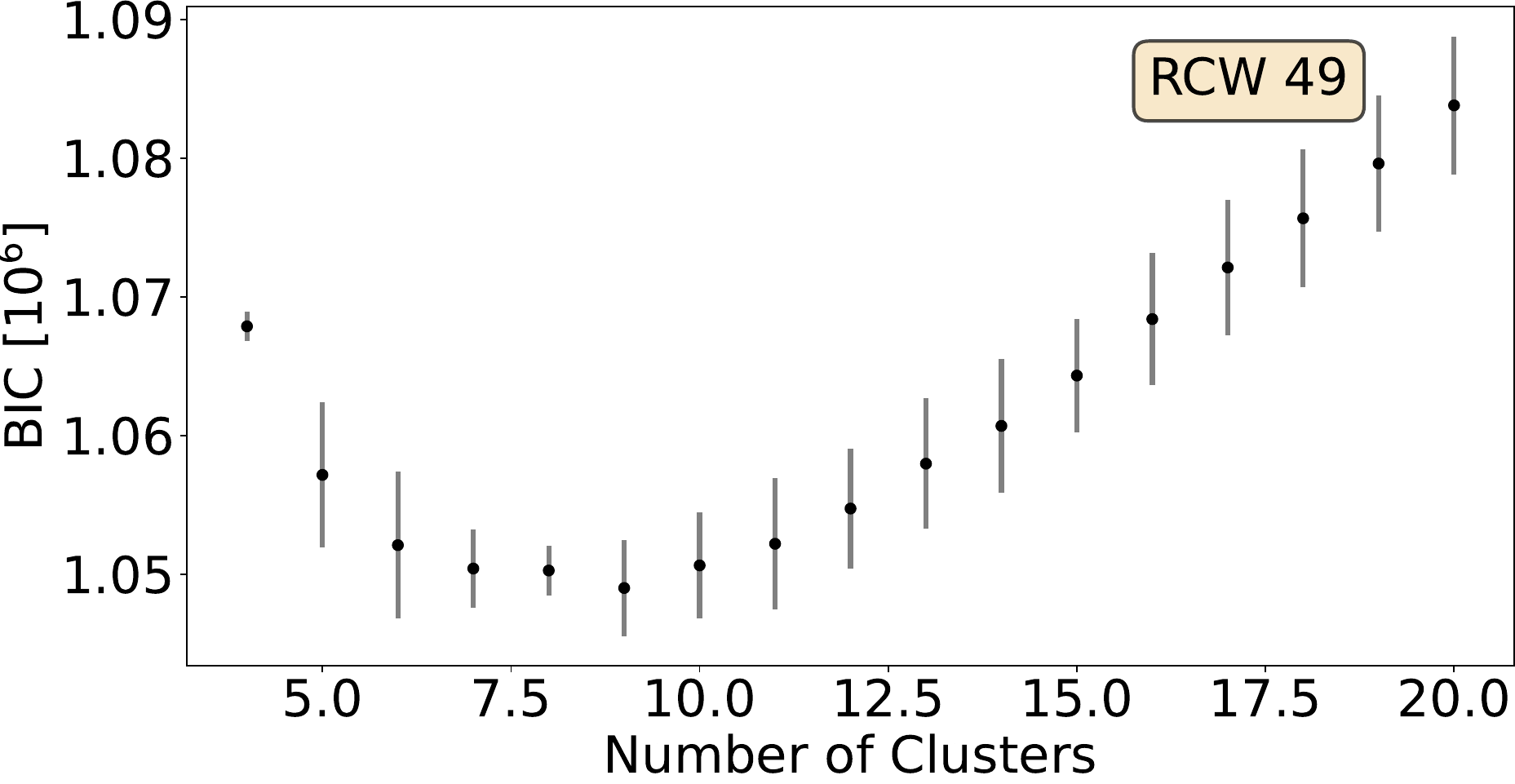}

\caption{Using the GMM described in \citet{Kabanovic2022}, BIC versus $n_{\rm input}$ plot (left panel) and domain map (right panel) for NGC~1977, RCW~120 and RCW~49.    
\label{fig:rcw120-slawa}}
\end{figure}

\begin{figure}[h]
\centering
\includegraphics[width=160mm]{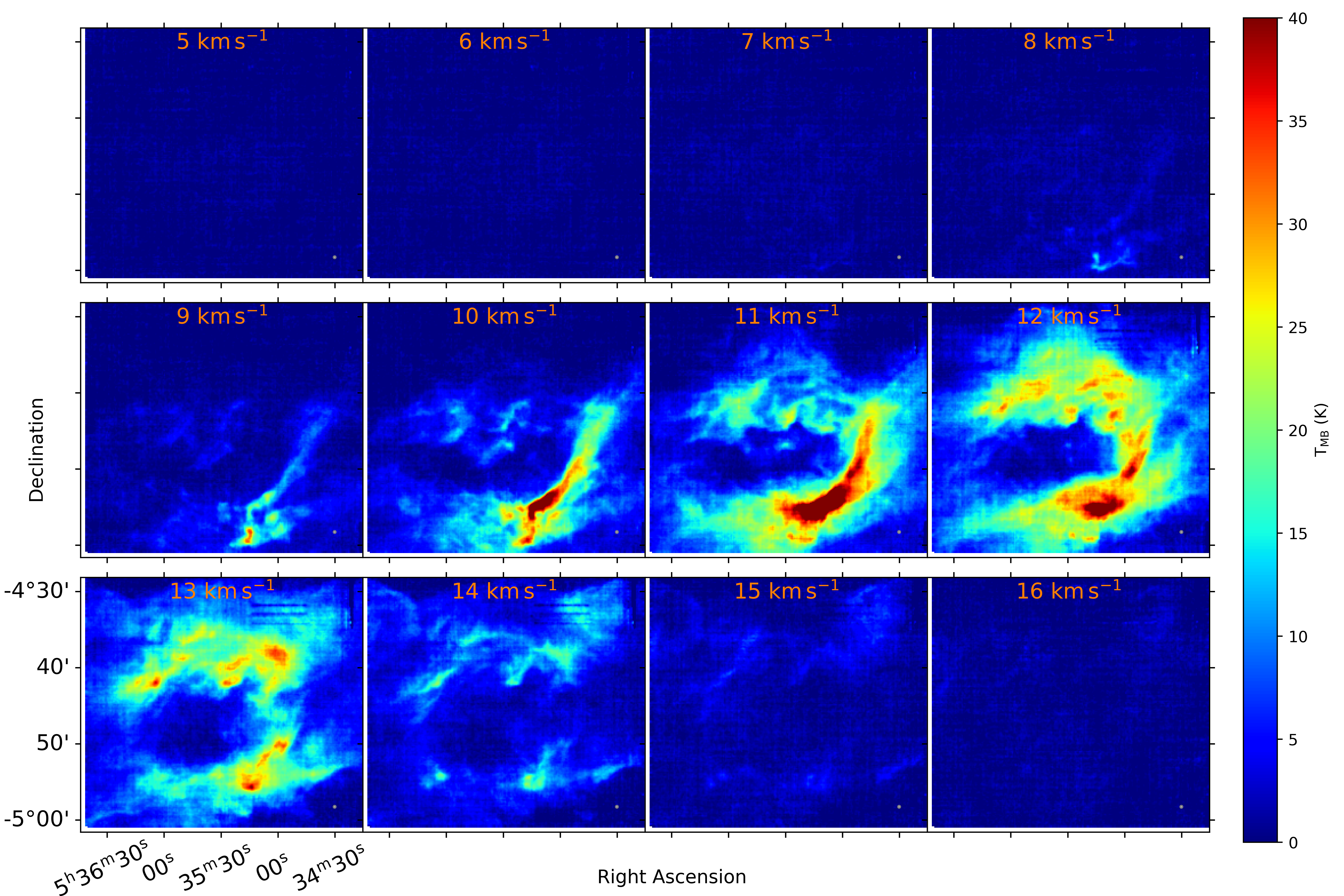}
\caption{Velocity channel maps of \cii\ emission towards NGC~1977. The beam size shown in the lower right.}
\label{fig:chan-maps-ngc1977}
\end{figure}

\begin{figure}[h]
\centering

\includegraphics[width=160mm]{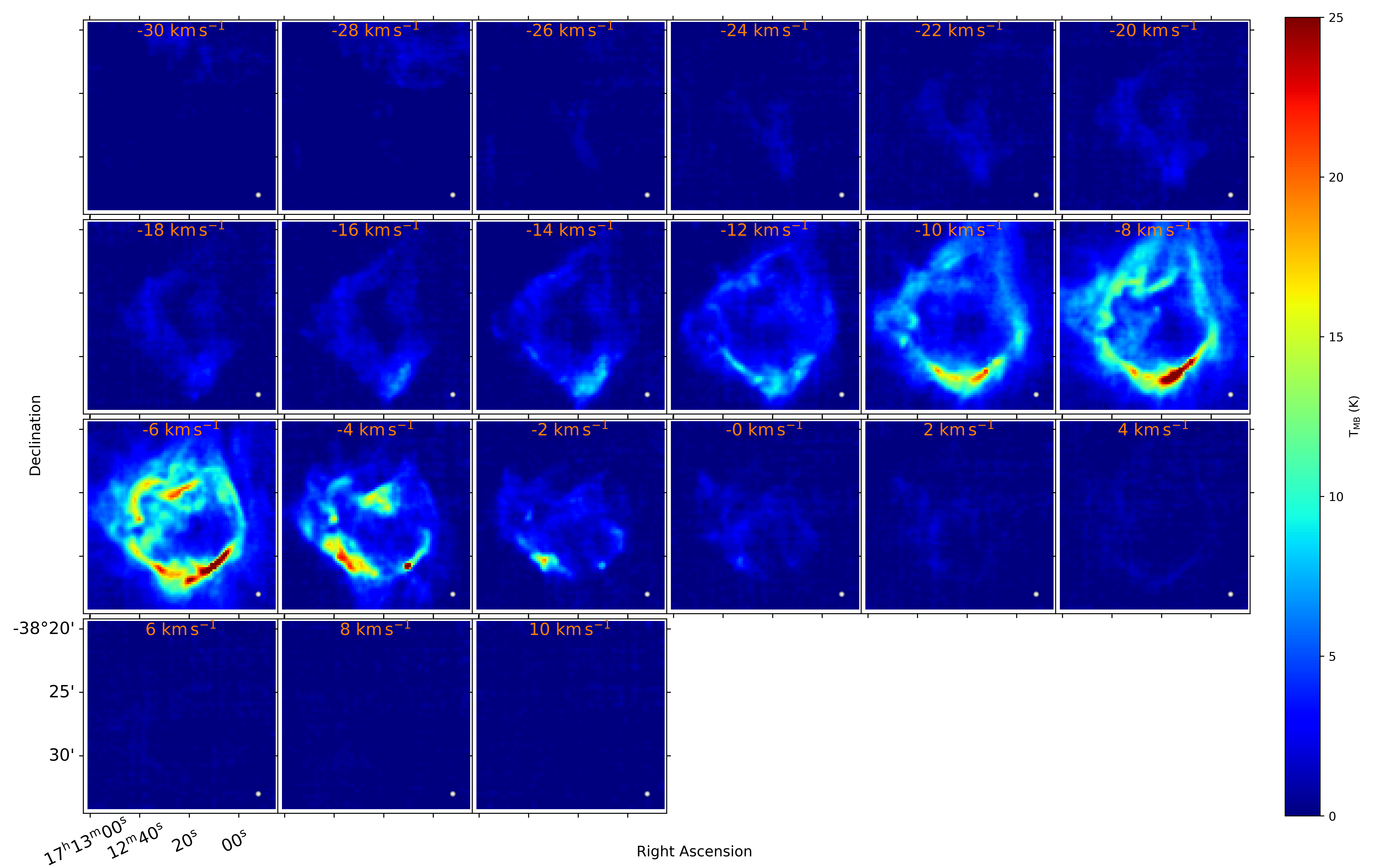}
\caption{Velocity channel maps of \cii\ emission towards RCW~120. The beam size shown in the lower right.}
\label{fig:chan-maps-rcw120}
\end{figure}

\begin{figure}[h]
\centering

\includegraphics[width=190mm]{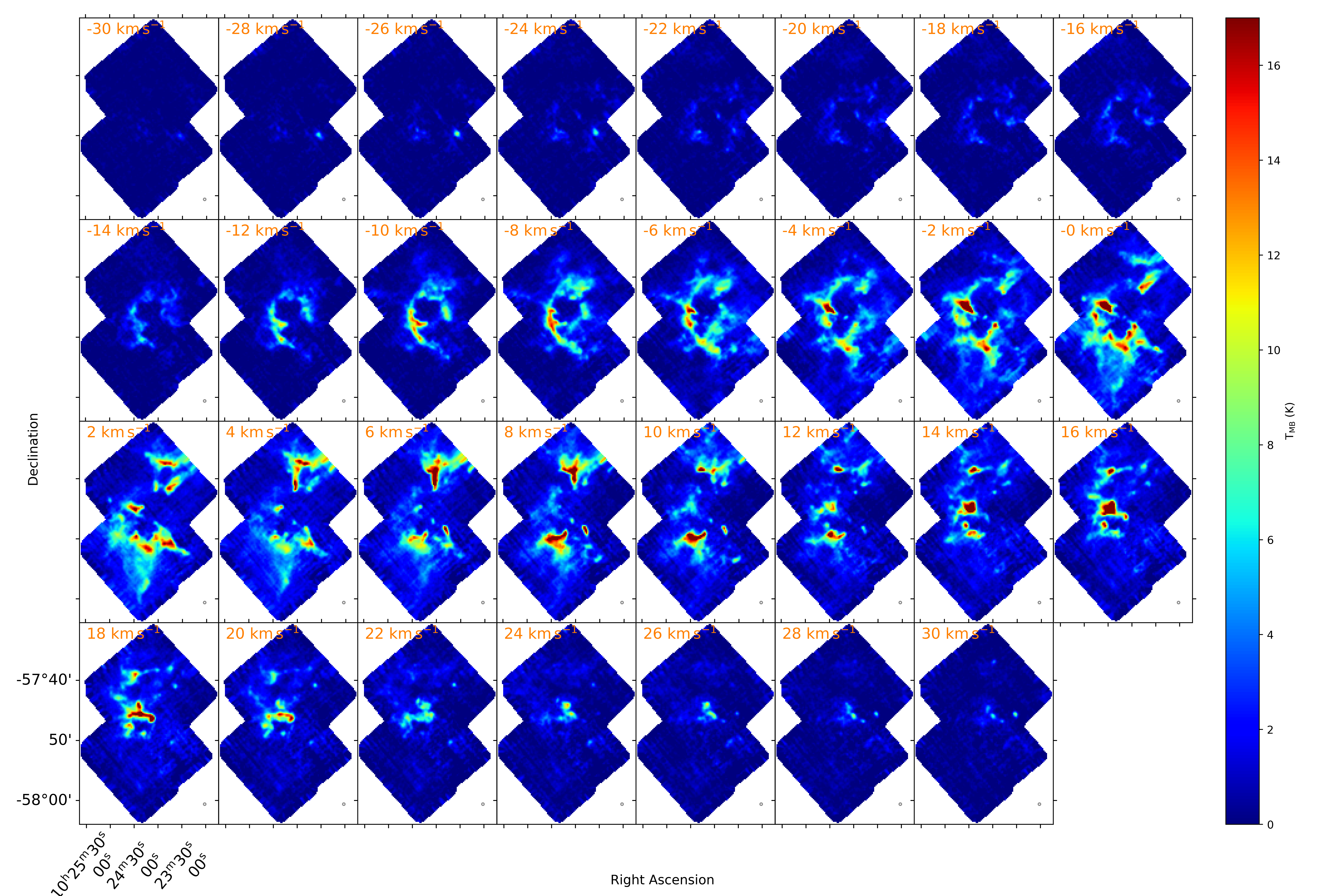}
\caption{Velocity channel maps of \cii\ emission towards RCW~49. The beam size shown in the lower right.}
\label{fig:chan-maps-rcw49}
\end{figure}

\begin{figure}[h]
\centering
\includegraphics[width=160mm]{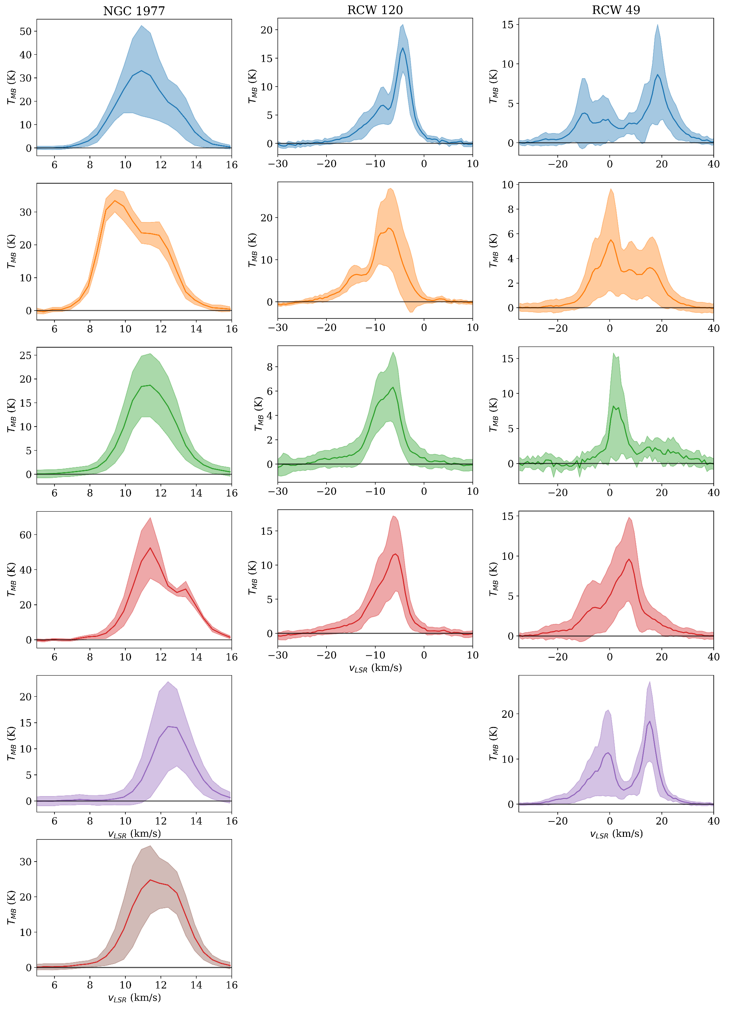}
\caption{Average spectra of all GMM identified clusters for NGC~1977 (left column), RCW~120 (middle column) and RCW~49 (right column). The dashed region represents the standard deviation depicting the variation in spectral profiles for all spectra per pixel corresponding to a cluster.  
\label{fig:ave-spec-std}}
\end{figure}

\begin{figure}[h]
\centering
\includegraphics[width=150mm]{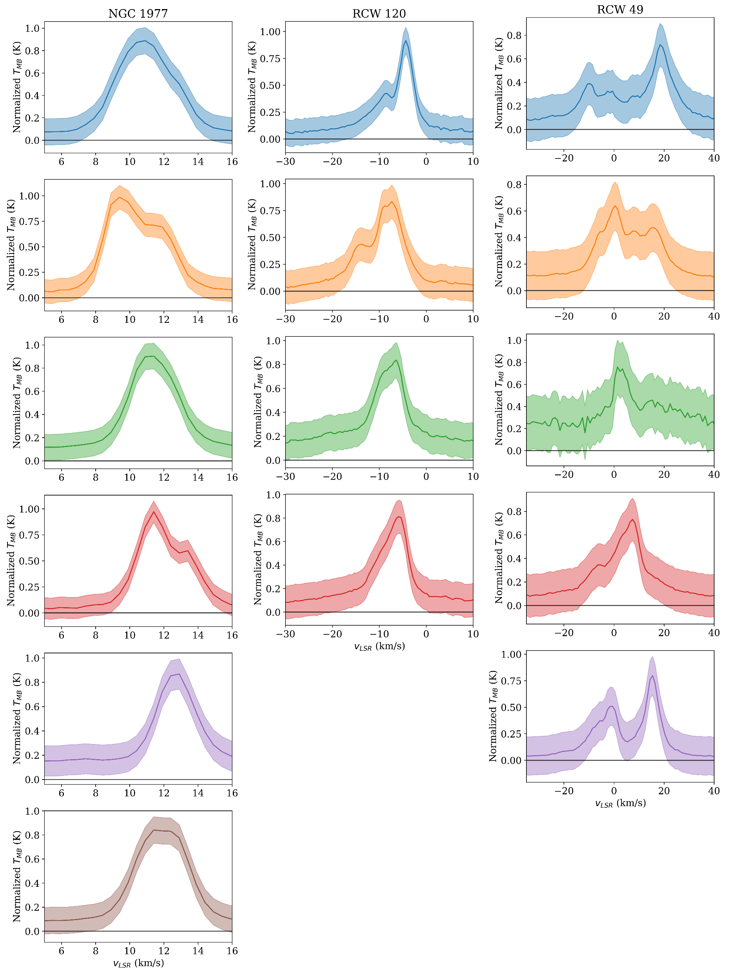}
\caption{Normalised average spectra of all GMM identified clusters for NGC~1977 (left column), RCW~120 (middle column) and RCW~49 (right column). The dashed region represents the standard deviation depicting the variation in spectral profiles for all spectra per pixel corresponding to a cluster.  
\label{fig:ave-spec-norm-std}}
\end{figure}

\section{Mean normalisation}\label{app:mean}




We tested the models using several normalisation techniques and found that the $mean$ normalization works better than the other techniques. 
For a single spectrum, the $mean$ normalisation is described as:

\begin{equation}
    T_{\rm mb}({\rm norm}) = T_{\rm mb} - T_{\rm mb}({\rm mean}) ({\rm K}),
\end{equation}

where $T_{\rm mb}$(norm) is the normalized spectral intensity per channel, $T_{\rm mb}$ is the observed main beam spectral intensity and $T_{\rm mb}(mean)$ is the mean of the main beam spectral intensities of all the spectra towards the entire source area. All temperatures are in K.

We ran the models with $mean$ normalisation on RCW~120 and 
Fig.~\ref{fig:rcw120-dm-mean} shows the domain map of RCW~120, where two clusters are identified (excluding the noise). The blue colored cluster depicts the broken (from the north and the east) shell of RCW~120 and the orange colored cluster comprises relatively brighter pixels in the region. Normalising the data set causes a decrease in the difference in intensities between various pixels. Thus, a lot of internal structure is lost in the $mean$ normalised results when compared to the domain map (Fig.~\ref{fig:rcw120-weights-dm-spec}) obtained without normalising the data. It is possible to achieve the same level of internal structure in the normalised data set, however, one needs to increase the input number of clusters. This will make the entire process slower when compared to running the models without using any normalisation.

\section{Traditional GMM}\label{app:slawa-models}

To investigate the difference between the `traditional' GMM and the models presented in this work, we ran the models described in \citet{Kabanovic2022} on NGC~1977, RCW~120 and RCW~49. 
Fig.~\ref{fig:rcw120-slawa}, bottom panel, shows the BIC for the three sources, with the minimum BIC found at $n_{\rm input}$ = 33, 6, 9 for NGC~1977, RCW~120 and RCW~49, respectively. These values match well with the $n_{\rm output}$ reported in Table~\ref{tab:input-parameters} for RCW~120 and RCW~49.



After excluding a single cluster that corresponds to only noise, the domain map of RCW~120 (Fig.~\ref{fig:rcw120-slawa}) is very similar to the one shown in Fig.~\ref{fig:rcw120-weights-dm-spec}. It shows clear signs of shell leakage in the north-west and in the east (breaking in the orange cluster). The models also identify the higher intensity pixels as an independent cluster (red). The blue cluster, however, encompasses pixels with low intensity (surrounding the shell shown in green).

In RCW~49, apart from the four (red, pink, brown and dark blue) 
clusters that correspond to the noise artefacts, the domain map (in Fig.~\ref{fig:rcw120-slawa}) identifies the major physical structures in RCW~49. 
Gray cluster corresponds to the broken shell. The green and orange 
clusters correspond to the northern and southern clouds. The light blue cluster corresponds to the ridge and also the shell in the north-east. However, unlike the results presented in Sect.~\ref{sec:GMM-results}, the traditional GMM models are unable to identify the pillar in RCW~49.     

Unlike RCW~120 and RCW~49, the minimum BIC in NGC~1977 occurs at a significantly high value leading to a domain map with 33 clusters.  
One of the possible reasons for the traditional models to converge at such a large number of clusters could be the large fluctuations in the spectral profiles along the shell which are more pronounced in NGC~1977 than in the other sources. 
Additionally, the relatively higher 
noisy pixels in the NGC~1977 data cause further artefacts in the generated domain map.
In comparison, the GMMis model results for NGC~1977 presented in Sect.~\ref{sec:GMM-results} deal better with these variations introduced by velocity gradients or noise. The GMMis model converges at 8 clusters for NGC~1977, clearly identifying the major physical structures in the source. Thus, emphasizing the potential in GMMis in analysing large data sets. 


\section{Velocity channel maps}\label{app:vel-chan-maps}

Figures~\ref{fig:chan-maps-ngc1977}, \ref{fig:chan-maps-rcw120} and \ref{fig:chan-maps-rcw49} show the velocity channel maps of \cii\ emission towards NGC~1977, RCW~120 and RCW~49. Similar maps have been previously presented in \citet{Pabst2022}, \citet{Kabanovic2022} and \citet{Tiwari2021}.

The shell of NGC~1977 can be seen in Figure~\ref{fig:chan-maps-ngc1977}. The southern part of the shell is most intense within 9 to 12~km~s$^{-1}$, while the northern part is most intense within 11 to 13~km~s$^{-1}$. This velocity gradient is the reason that the northern and southern parts of the shell are identified as separate clusters by the models (Fig.~\ref{fig:ngc1977-wieights-dm-spec}). 
In RCW~120, the \cii\ velocity channel maps (Fig.~\ref{fig:chan-maps-rcw120}) illustrate the ring dynamics from -12 to -2~km~s$^{-1}$. The break in the north-western break in the shell is also visible in the channel maps.
The different structures of RCW~49 can be seen in the velocity channel maps (Fig.~\ref{fig:chan-maps-rcw49}). 
The broken expanding shell is most intense from -14 to 0~km~s$^{-1}$. 
The northern and southern clouds are most intense from 2 to 8~km~s$^{-1}$. The pillar is brightest within 4 to 12~km~s$^{-1}$, while the ridge is brightest within 16 to 22~km~s$^{-1}$. The domain map of RCW~49 (Fig.~\ref{fig:rcw49-weights-dm-spec}) identifies all these structures as independent clusters except the pillar, which is part of the southern cloud.

\section{Average spectrum of each cluster}\label{app:avg-spec-std}

Figure~\ref{fig:ave-spec-std} shows the average spectra of all clusters identified by GMM for NGC~1977, RCW~120 and RCW~49. It can be seen that all pixels assigned to a given cluster have similar spectral profiles. Most of the average spectra have relatively small dispersion which means that all the spectra corresponding to that cluster have very similar intensities and line shapes. The relatively larger dispersion seen in some of the average spectra is mainly because of the difference in the line intensity, while the line profiles corresponding to all the pixels associated with one cluster are similar. To illustrate it, we show the normalised average spectra of all clusters in Fig.~\ref{fig:ave-spec-norm-std} and it can be seen that the dispersion is overall less in this case. 
Thus, the models are able to classify spectra based on their line profiles very well, further validating their accuracy.




\section{The case of cluster assignment with low probability}\label{app:low-prob}

\begin{figure}[h]
\centering
\includegraphics[width=85mm]{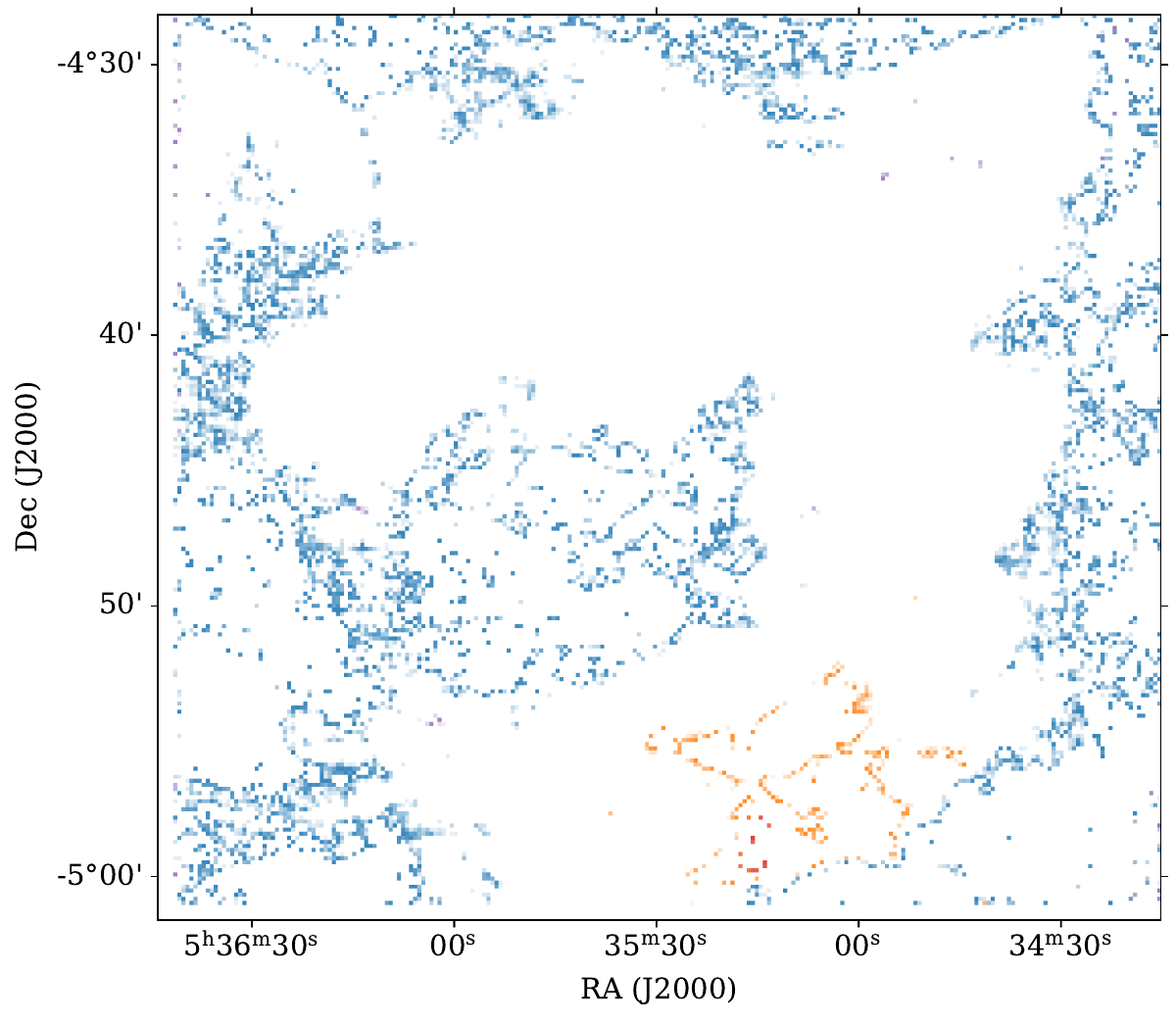}
\caption{Domain map of NGC~1977 from an identical model to the one presented in Fig.~\ref{fig:ngc1977-wieights-dm-spec}, however instead of assigning each pixel to the cluster to which it has the largest probability of belonging, we only plot points with a probability between 0.1 and 0.9.  
The color opacity of each pixel is proportional to this probability.}
\label{fig:ngc1977-low-prob}
\end{figure}

Figure~\ref{fig:ngc1977-low-prob} highlights the pixels in NGC~1977 domain map that have low probabilities ($<$0.9) of getting assigned to their respective clusters as shown in Fig.~\ref{fig:ngc1977-wieights-dm-spec}. This is an example to understand the spatial distribution of these pixels. 
Firstly, we found that the map contains relatively few of these pixels. Also, they are more localised to the boundaries of the identified clusters implying a more gradual transition from one cluster to another in contrast to what is seen in the domain maps themselves.

\end{document}